\documentclass[trackchanges,twocolumn]{aastex701}
\usepackage{tabularx}
\usepackage{xcolor}
\usepackage{colortbl}
\usepackage{array}
\usepackage{caption}

\def\mjysr{$\mathrm{MJy\,sr^{-1}}$}
\def\Wcmumsr{$\mathrm{W\,cm^{-2}\,\mu m^{-1}\,sr^{-1}}$}

\begin{document}

\title{Reliability of Zodiacal Background Models at Small Solar Elongations}

\correspondingauthor{Csaba Kiss (pkisscs@konkoly.hu)}
\author[0000-0002-8722-6875]{Csaba Kiss}
\affiliation{HUN-REN CSFK Konkoly Observatory, MTA Centre of Excellence, Konkoly Thege M. \'ut 15-17, Budapest, 1121, Hungary}
\affiliation{ELTE Eötvös Loránd University, Institute of Physics and Astronomy, P\'azm\'any P\'eter S\'et\'any 1/A, Budapest, 1117, Hungary}
\email{pkisscs@konkoly.hu}

\author[0000-0000-0000-0000]{Thomas G. Müller}
\affiliation{Max-Planck-Institut für extraterrestrische Physik, Garching, Germany}
\email{tmueller@mpe.mpg.de}

\author[0000-0002-9214-337X]{Javier Licandro}
\affiliation{Instituto de Astrofísica de Canarias (IAC), 38205 La Laguna, Tenerife, Spain}
\affiliation{Departamento de Astrofísica, Universidad de La Laguna, 38206 La Laguna, Tenerife, Spain}
\email{jlicandr@iac.es}

\author[0000-0002-6710-8476]{Luca Conversi}
\affiliation{European Space Agency -- ESRIN, Via Galileo Galilei, 00044 Frascati (RM), Italy}
\email{luca.conversi@esa.int}

\author[0000-0000-0000-0000]{Marco Delbo}
\affiliation{Universit\'e C\^ote d’Azur, Observatoire de la C\^ote d’Azur, CNRS, Laboratoire Lagrange, CS34229, 06304 Nice Cedex 4, France}
\affiliation{School of Physics and Astronomy, University of Leicester, Leicester, United Kingdom}
\email{marco.delbo@oca.eu}

\author[0000-0001-8058-2642]{Karri Muinonen}
\affiliation{Department of Physics, University of Helsinki, Gustaf H\"allstr\"omin katu 2, FI-00560 Helsinki, Finland}
\email{karri.muinonen@helsinki.fi}

\author[0000-0001-8585-204X]{Marcel Popescu}
\affiliation{Institute of Space Science (ISS) - INFLPR subsidiary, Ilfov \& 077125, Romania}
\email{marcel.popescu@spacescience.ro}

\author[0000-0000-0000-0000]{Paolo Tanga}
\affiliation{Universit\'e C\^ote d’Azur, Observatoire de la C\^ote d’Azur, CNRS, Laboratoire Lagrange, CS34229, 06304 Nice Cedex 4, France}
\email{paolo.tanga@oca.eu}

\author[0000-0002-1698-605X]{Róbert Szakáts}
\affiliation{HUN-REN CSFK Konkoly Observatory, MTA Centre of Excellence, Konkoly Thege M. \'ut 15-17, Budapest, 1121, Hungary}
\email{szakats.robert@csfk.org}

\begin{abstract}

Present models of the zodiacal light (scattered sunlight from interplanetary dust in the visible) and zodiacal emission (thermal emission in the mid-infrared) are primarily constrained by surface-brightness measurements obtained at relatively large solar elongations ($\epsilon \gtrsim 60^\circ$). As a result, their predictive accuracy at smaller elongations remains uncertain. This is particularly relevant for solar elongations of $\epsilon \approx 30^\circ$–$60^\circ$, where next-generation near-Earth-object discovery missions such as NEO Surveyor and NEOMIR are designed to operate.
To evaluate the reliability of these models at small solar elongations, we compile and reanalyze the available 
small-elongation zodiacal-light and zodiacal-emission measurements and provide the first systematic validation of modern zodiacal-light models in a regime where they have never been independently tested. 
{We compare two models derived from the COBE/DIRBE survey: one constrained primarily by the spatial and temporal variation of the monochromatic sky brightness, and another that additionally uses interband color information and fixes the absolute normalization by requiring the minimum high-latitude residual at $25\,\mu{\rm m}$ to vanish. We find that the former reproduces the observations significantly better. 
At wavelengths most relevant to infrared near-Earth-object surveys ($\sim8$--$10\,\mu{\rm m}$), it typically agrees with measurements to within $\sim10\%$ at $\epsilon \approx 40^\circ$--$60^\circ$. Although it also performs better at $\epsilon \approx 30^\circ$, significant discrepancies remain.}
These results show that zodiacal emission at small solar elongations is still only weakly constrained and that its absolute brightness remains uncertain. New mid-infrared observations are required to improve background estimates for future infrared survey missions.

\end{abstract}

\submitjournal{PASP}


\section{Introduction}

{  Zodiacal light and zodiacal emission arise from sunlight scattered and
thermally re-emitted by interplanetary dust in the inner Solar System.
The zodiacal cloud is a flattened, slightly asymmetric distribution
replenished predominantly by Jupiter-family comets and, to a lesser
extent, asteroid collisions \citep{Nesvorny2010ApJ...713..816N}, and
shaped by Poynting--Robertson drag, collisions, radiation pressure, and
planetary perturbations
\citep{Lasue2020P&SS..19004973L,Rigley2022MNRAS.510..834R}.
This dust dominates much of the optical and infrared diffuse sky, particularly at low ecliptic latitudes and small solar elongations \citep{Leinert1998}. It is therefore a major determinant of the sensitivity and survey efficiency of infrared Near-Earth Object (NEO) missions such as NEO Surveyor \citep{Mainzer2023PSJ.....4..224M} and NEOMIR \citep{2024SPIE13092E..2HC,Conversi2026NEOMIR}. NEO Surveyor is designed to observe over
$45\degr\leq\epsilon\leq120\degr$ and $|\beta| \leq40\degr$, while NEOMIR will operate over $30\degr\leq\epsilon\leq70\degr$, where the
lines of sight encounter enhanced dust densities and temperatures.

Interplanetary-dust emission peaks in the mid-infrared, overlapping the
$\sim4$--$10\,\mu$m range favorable for detecting NEO thermal emission \citep{Mainzer2023PSJ.....4..224M,Masiero2024PSJ.....5..113M,Masiero2024PSJ.....5..222M}. At these wavelengths it generally exceeds Galactic cirrus and extragalactic backgrounds by orders of magnitude and sets the principal diffuse photon-noise floor \citep{Leinert1998}. Its uncertainty consequently propagates into estimates of limiting sensitivity, integration time, survey completeness, and discovery yield. The high background may also limit exposure times and detector dynamic range or cause non-linearity and saturation.

These effects are especially important at
$\epsilon\lesssim60\degr$, and particularly near $30\degr$, where the dust temperature and column density increase rapidly along the line of sight \citep{Murdock1985AJ.....90..375M,Price2003}. Existing models, constrained largely near $90\degr$, must be extrapolated into a regime
where the dust distribution and optical properties are less certain. Accurate modelling is therefore required to maintain uniform survey sensitivity and to constrain predicted discovery rates
and systematic uncertainties in NEO population studies
\citep{Mainzer2015Surveyor,Conversi2026NEOMIR}.

To assess the reliability of current zodiacal background predictions at small solar elongations, in this paper we compare zodiacal-background models with archival measurements from spacecraft and sounding rockets, including observations obtained from within Earth's orbit. Section~\ref{sect:zlmodels} summarizes the relevant models, Section~\ref{sect:zlcomp} compares the principal COBE-based
implementations, and Section~\ref{sect:zlmeas} tests their predictions against measurements from Helios, MSX, and the ZIP rocket experiment.
Our conclusions are presented in Section~\ref{sect:conclusions}.}

\section{Zodiacal Emission Models Based on COBE/DIRBE \label{sect:zlmodels}}

The most widely used empirical models of zodiacal emission in the infrared remain those derived from the \textit{COBE}/DIRBE dataset, most notably the models of \citet{Kelsall1998} and \citet{Wright1998}. Both models were constructed using the same underlying observational dataset, namely the ten-band (1.25--240~$\mu$m) all-sky measurements obtained by DIRBE over approximately ten months, but differ substantially in their fitting philosophy, assumptions, and resulting absolute normalization of the zodiacal light.

DIRBE’s observational strategy, which repeatedly scanned the full sky at solar elongations near $90^\circ$ while the Earth orbited the Sun, enabled the separation of time-variable zodiacal emission from relatively static Galactic and extragalactic backgrounds. The seasonal modulation of brightness as a function of viewing geometry provided the principal constraint on the three-dimensional distribution of interplanetary dust (IPD).

\subsection{The Kelsall et al.\ (1998) Model}

The model developed by \citet{Kelsall1998} is a parametric, physically motivated representation of the IPD cloud. It consists of several distinct components: (1) a smooth cloud with a radially decreasing density and vertical flattening, (2) three pairs of asteroidal dust bands associated
with major asteroid families, (3) a circumsolar ring near 1~au, and (4) an Earth-trailing overdensity or ``blob.'' Each component is described by an analytic density distribution, together with prescriptions for scattered sunlight and thermal emission.

{The geometrical parameters of the dust distribution were constrained using the seasonal variation of the DIRBE sky brightness. Although the solar spectrum and the thermal blackbody kernel supplied a common underlying spectral form, the fit included independent emissivity or albedo scale
factors for the individual DIRBE bands. The ten wavelengths were therefore radiometrically only weakly coupled, providing a deliberately conservative fit that did not rely on the DIRBE absolute calibration scale. Moreover, the fitting statistic was based on the temporal variation about the
mission-averaged brightness rather than on the absolute sky level, leaving the isotropic zero point unconstrained. Consequently, the original model is defined strictly at the ten DIRBE wavelengths; predictions at intermediate wavelengths require an additional interpolation prescription. Emission remaining after zodiacal subtraction was interpreted as a combination of Galactic foregrounds and an isotropic extragalactic background.}

\subsection{The Wright (1998) Model}

\citet{Wright1998} introduced an alternative zodiacal model derived from the same DIRBE dataset but incorporating an additional constraint known as the ``strong no-zodi principle.'' This assumption enforces that at 25~$\mu$m the minimum residual at high Galactic latitude after subtraction of a zodiacal light model from the DIRBE observations has to be zero.
The practical effect of this assumption is a renormalization of the zodiacal emission to higher absolute levels than in the Kelsall model, particularly in the mid-infrared. While the spatial structure of the IPD cloud in the Wright model remains broadly similar to that of \citet{Kelsall1998}, the resulting zodiacal intensities can differ by tens of percent.
Unlike the Kelsall model, the Wright model does not include a separate Earth-trailing dust cloud component; attempts to fit such a component yielded an unphysical negative density, suggesting that its contribution is instead absorbed implicitly into the fitted smooth-cloud brightness distribution.
This difference propagates into estimates of the cosmic infrared background and into downstream applications such as sky background prediction tools.
These differences also suggest that the Wright model should primarily be regarded as an empirical representation of the DIRBE observations. While this approach provides an appropriate description within the range of the observational data, its extrapolation to observing geometries that are not constrained directly by DIRBE measurements may be less reliable than models based on a more explicit physical decomposition of the dust cloud.

\subsection{Relationship to IRAS and Asteroidal Dust Bands}

Although the existence of asteroidal dust bands was first established using \textit{IRAS} data \citep{Low1984}, the Kelsall and Wright models were not fit directly to IRAS observations. IRAS data lacked the temporal sampling and absolute calibration stability required for seasonal fitting of the zodiacal cloud. Instead, DIRBE’s much coarser angular resolution limited the zodiacal band modeling to the three most prominent band pairs detectable in the DIRBE maps.

\citet{2013MNRAS.429.2894R} developed an improved model of zodiacal infrared emission by jointly fitting \textit{IRAS} and \textit{COBE}/DIRBE data, decomposing the interplanetary dust cloud into cometary, asteroidal, and interstellar components. They found that cometary dust dominates the infrared emission within the inner Solar System, with smaller but significant contributions from asteroidal bands and an approximately isotropic interstellar component.

\subsection{Comparison with Other Space-Based Datasets}

Neither the Kelsall nor Wright models were originally fit to datasets beyond DIRBE. However, both have been extensively compared with observations from later space missions. Mid-infrared measurements from \textit{Spitzer} revealed systematic residuals at the few-percent level relative to the Kelsall model, suggesting additional structure or slight warping of the IPD cloud {\citep{Krick2012}. }

\textit{AKARI} all-sky surveys at 9 and 18~$\mu$m provided higher-resolution views of the zodiacal dust bands and showed discrepancies with the DIRBE-based band parameters, particularly in band latitude and contrast \citep{Tsumura2013}. \textit{Planck} observations extended the DIRBE zodiacal components into the submillimeter regime, fitting emissivities for the same structural components rather than redefining the dust distribution itself \citep{Planck2014Zodi}.

Optical and near-infrared comparisons using \textit{HST} and ground-based data indicate that extrapolations of the DIRBE-fitted models into the scattered-light regime can lead to systematic offsets, underscoring the wavelength-limited nature of the original fits \citep{Arendt2003ApJ...585..305A}.

 \citet{OBrien2026ApJ..1000....6O} presented a new empirical model of zodiacal light at optical and near-infrared wavelengths, derived from more than 5{,}000 sky surface-brightness measurements obtained with the \textit{Hubble Space Telescope} as part of the SKYSURF program. Their model, termed \textit{ZodiSURF}, builds on the classical DIRBE-based framework by directly constraining the dust scattering phase function and wavelength-dependent albedo using in situ optical data, achieving significantly improved fits to observed zodiacal light at elongations $\gtrsim 80^\circ$. The analysis also revealed a residual diffuse component not captured by standard interplanetary dust models, which the authors interpret as possible evidence for an additional, approximately spherical dust population distinct from the canonical zodiacal cloud {, also discussed by the CIRER team \citep{Korngut2022}. }

 {The Cosmic Infrared Background ExpeRiment (CIBER) was a sounding-rocket payload that measured the diffuse sky at approximately $0.75$--$2.1\,\mu{\rm m}$. CIBER detected near-infrared fluctuations stronger than expected from known galaxies and stars, potentially associated with diffuse intrahalo light, while its absolute-spectrum measurements found excess emission whose interpretation remains limited by zodiacal-light subtraction \citep{Zemcov2014,Matsuura2017}. Its zodiacal-light observations revealed a red solar-scattered spectrum with a possible $0.9\,\mu{\rm m}$ silicate feature \citep{Tsumura2010}; Fraunhofer-line spectroscopy suggested an additional weakly modulated component of $46\pm19\,{\rm nW\,m^{-2}\,sr^{-1}}$ at $1.25\,\mu{\rm m}$ \citep{Korngut2022}; and spectropolarimetry favored scattering by relatively large, absorbing dust grains \citep{Takimoto2022}. The polarization fields covered solar elongations of approximately $90^\circ$--$132^\circ$, so CIBER did not probe the inner zodiacal cloud at elongations below $90^\circ$.}

\subsection{Solar Elongation Coverage and Limitations}

A critical limitation of the DIRBE-based zodiacal models is their restricted solar elongation coverage. DIRBE observed the sky primarily at elongations near $90^\circ$, with only modest excursions away from this geometry. As a result, the Kelsall and Wright models are directly constrained only for elongations roughly in the range $60^\circ \lesssim \epsilon \lesssim 120^\circ$.

Comparisons with other datasets that probe smaller elongations have generally relied on extrapolations of the DIRBE-fitted models rather than independent constraints. \textit{Spitzer} observations, which can reach elongations of $\sim$85$^\circ$, do not probe significantly closer to the Sun \citep{Krick2012}. In contrast, heliospheric imagers and inner-solar-system instruments capable of observing at elongations $\lesssim30^\circ$ have demonstrated deviations from simple extrapolations of the DIRBE-based models, particularly in the radial dust density profile and forward-scattering behavior \citep[e.g.,][]{Leinert1998}.

To date, no widely adopted zodiacal emission model constrained by absolute photometry at solar elongations below $\sim60^\circ$ exists that is directly compatible with the DIRBE-based framework. This gap is particularly relevant for studies of zodiacal emission at small elongations, where line-of-sight integration effects and dust populations interior to 1~au become increasingly important.

\subsection{IPAC Background Model Implementations}

A zodiacal light/emission prediction tool widely used by the community is the IPAC Background Model\footnote{https://irsa.ipac.caltech.edu/applications/BackgroundModel/}. Background Model Version~1 adopts a combination of the \citet{Kelsall1998} and \citet{Reach1997Icar..127..461R} parametrizations of the zodiacal dust cloud and the asteroid bands {\citep{ipacv1}}, while Version~4 implements a \citet{Wright1998}-based zodiacal model following the refinements of \citet{Gorjian2000}. Although both versions are ultimately derived from the same DIRBE dataset, they differ in absolute zodiacal intensity and therefore in predicted infrared background levels. These models further include Galactic and extragalactic components not present in the original zodiacal emission fits. Besides an Earth-based observer, the IPAC Background Model also supports predictions for the Sun--Earth L2 Lagrange point by adopting the James Webb Space Telescope ephemeris over the interval 2018 October 1 to 2029 April 30.

\subsection{ZodiPy \label{sect:zodipy}}

{ZodiPy is an open-source, Astropy-affiliated Python package that evaluates published parametric models of zodiacal scattered light and thermal emission \citep{San2022,San2024}. It computes the surface brightness by integrating the three-dimensional IPD density and emission model along a specified line of sight, accounting for the observing time, pointing direction, observer position, and either a monochromatic wavelength or an instrumental bandpass. Its principal advantage here is the ability to evaluate the same model from arbitrary locations in the Solar System, including the Sun--Earth L1 region and the heliocentric positions of the Helios spacecraft. ZodiPy implements the \citet{Kelsall1998} DIRBE model over 1.25--240\,$\mu$m, as well as its Planck extension at microwave frequencies; its DIRBE implementation was validated by comparison with timestreams generated by the original DIRBE Zodiacal Light Prediction Software \citep{San2022}.
Comparisons between ZodiPy and the Kelsall-based IPAC V1 estimator test the consistency of two implementations of the same underlying model. ZodiPy additionally enables predictions from observer positions unsupported by the IPAC interface.}

\section{Comparison of DIRBE-based models at small solar elongations \label{sect:zlcomp}}

\begin{figure}[ht!]
    \centering
    \includegraphics[width=0.99\linewidth]{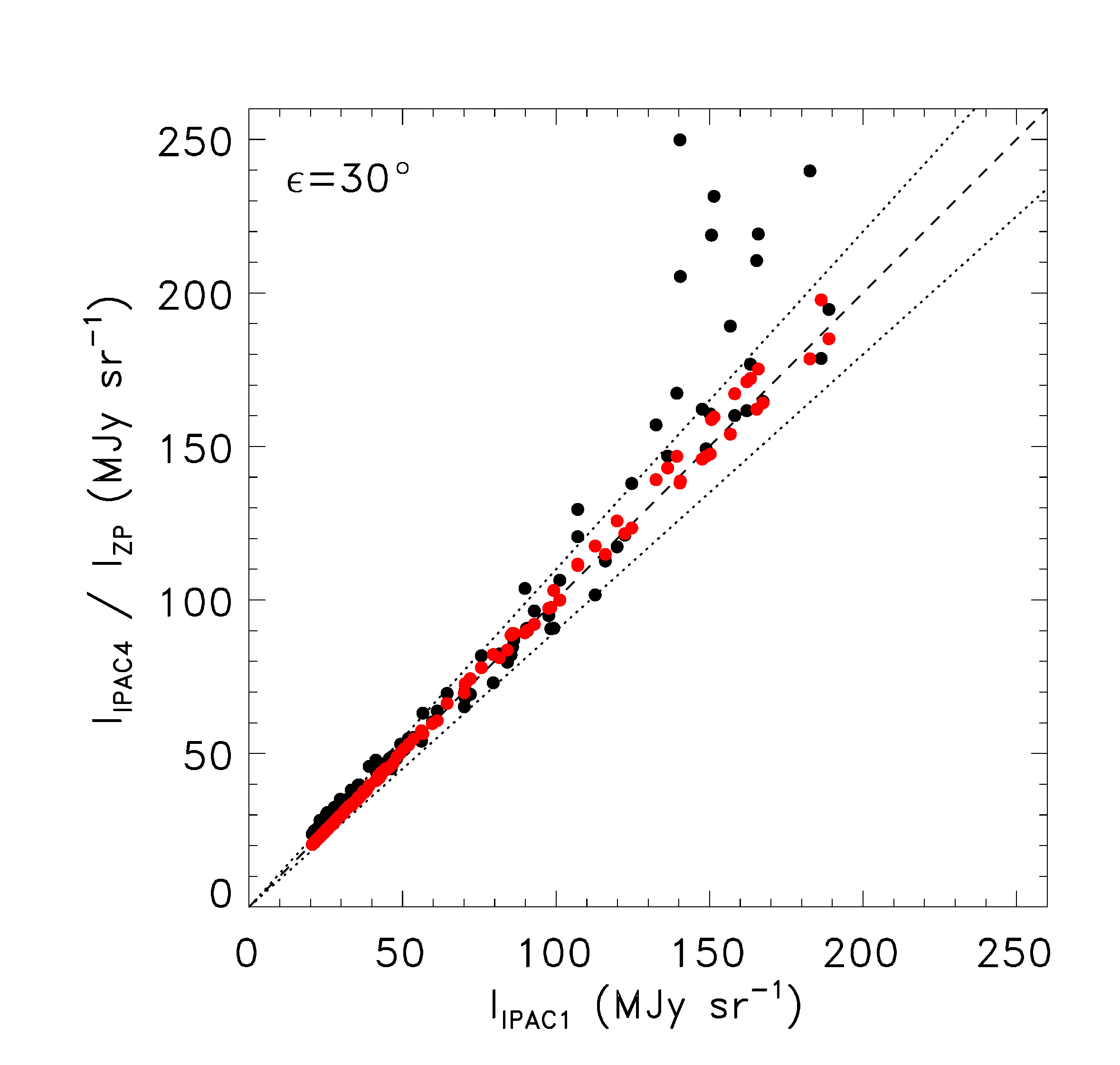}
    \includegraphics[width=0.99\linewidth]{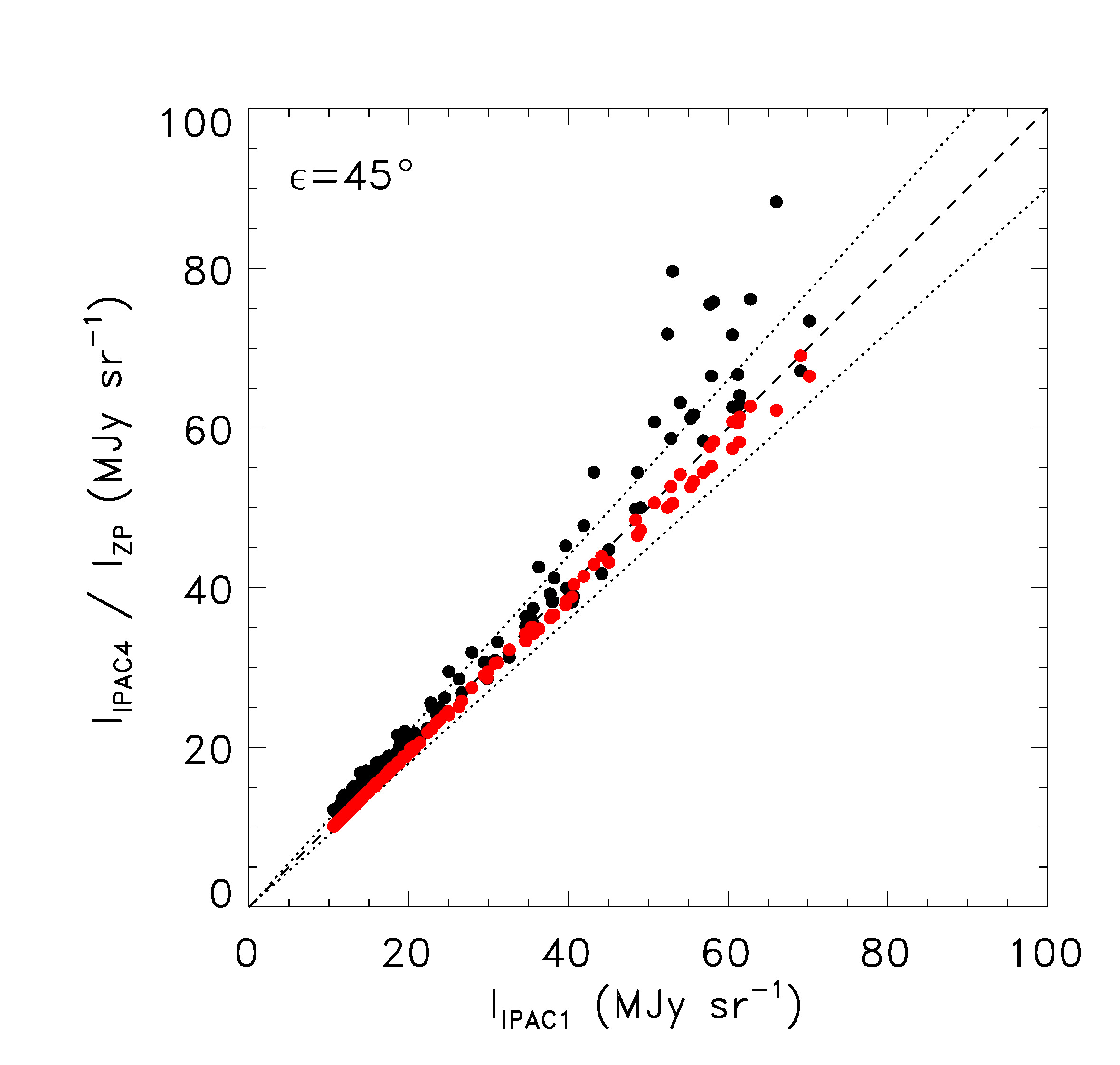}    
    \caption{Comparison of DIRBE-based model predictions at solar elongations of 30\degr\ and 45\degr, at 8\,$\mu$m. Surface brightnesses predicted by the IPAC V4 model (black points) and ZodiPy (red points) are plotted against the corresponding IPAC V1 predictions. High values correspond to low ecliptic latitudes. The dashed line denotes the 1:1 relation, while the dotted lines indicate deviations of $\pm$10\%.}
    \label{fig:sfb}
\end{figure}

As the DIRBE-based model implementations described above are widely used to estimate near- and mid-infrared sky backgrounds for the design and performance assessment of infrared space missions, it is important to understand how their predictions diverge at small solar elongations. As a simple illustration, we compared predictions from IPAC V1, IPAC V4, and ZodiPy at 8\,$\mu$m, a wavelength relevant to both NEOMIR and NEO Surveyor, for observations at a solar elongation of 30\degr\ and 45\degr\ throughout the year. 

{The DIRBE observations used to construct the models considered here covered solar elongations of $64^\circ \leq \epsilon \leq 124^\circ$. The predictions examined in this section extend to $\epsilon=30^\circ$ and therefore represent extrapolations beyond the range directly constrained by the fitted data. The purpose of this comparison is to quantify how the different DIRBE-based implementations diverge when applied to the small-elongation observing geometries relevant to NEO Surveyor and NEOMIR; it should not be interpreted as an observational validation of any model in this
regime. Such validation using independent small-elongation
measurements is presented in Sect.~\ref{sect:zlmeas}.}

\begin{figure}
    \centering
    \includegraphics[width=0.99\linewidth]{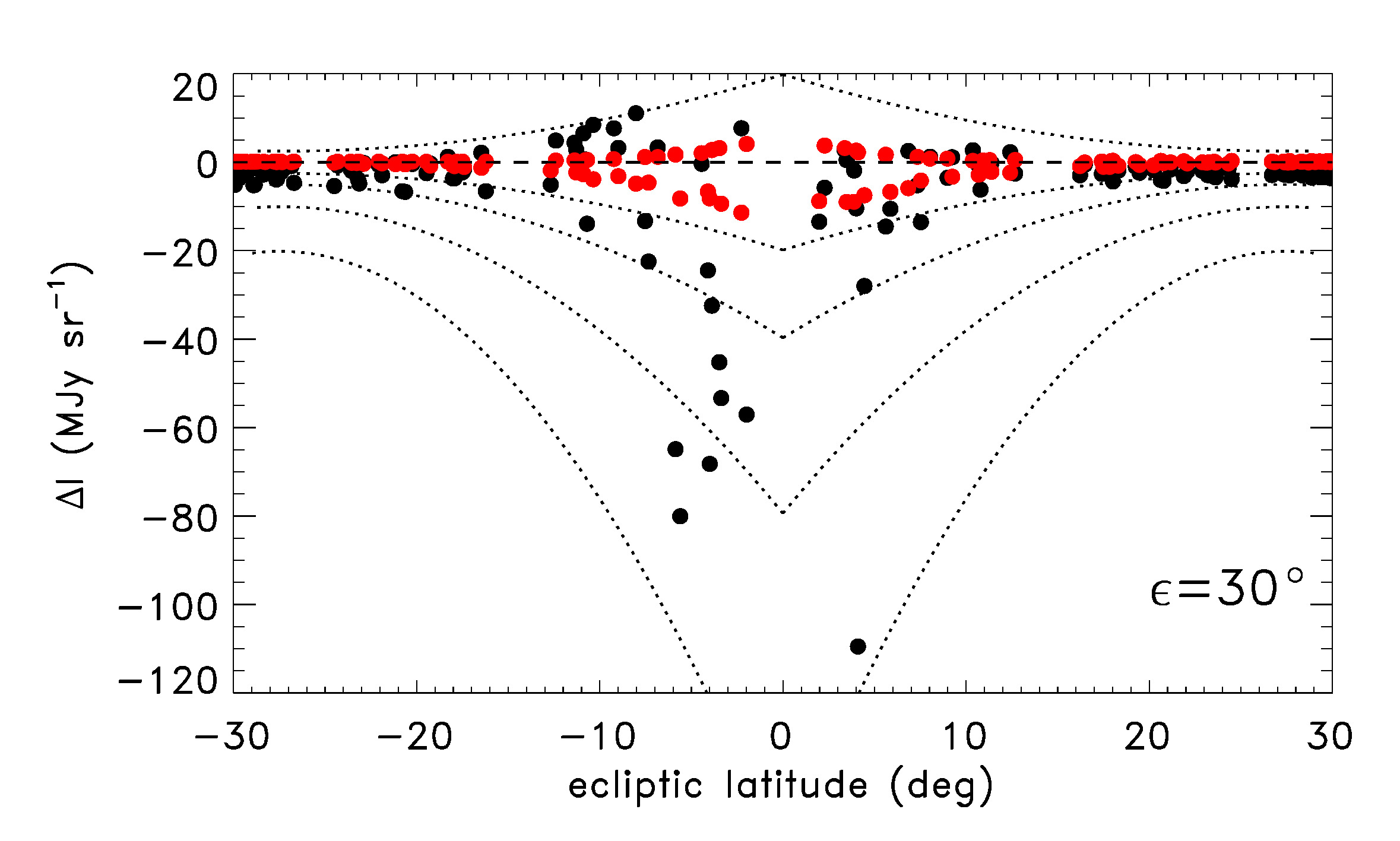}
       \includegraphics[width=0.99\linewidth]{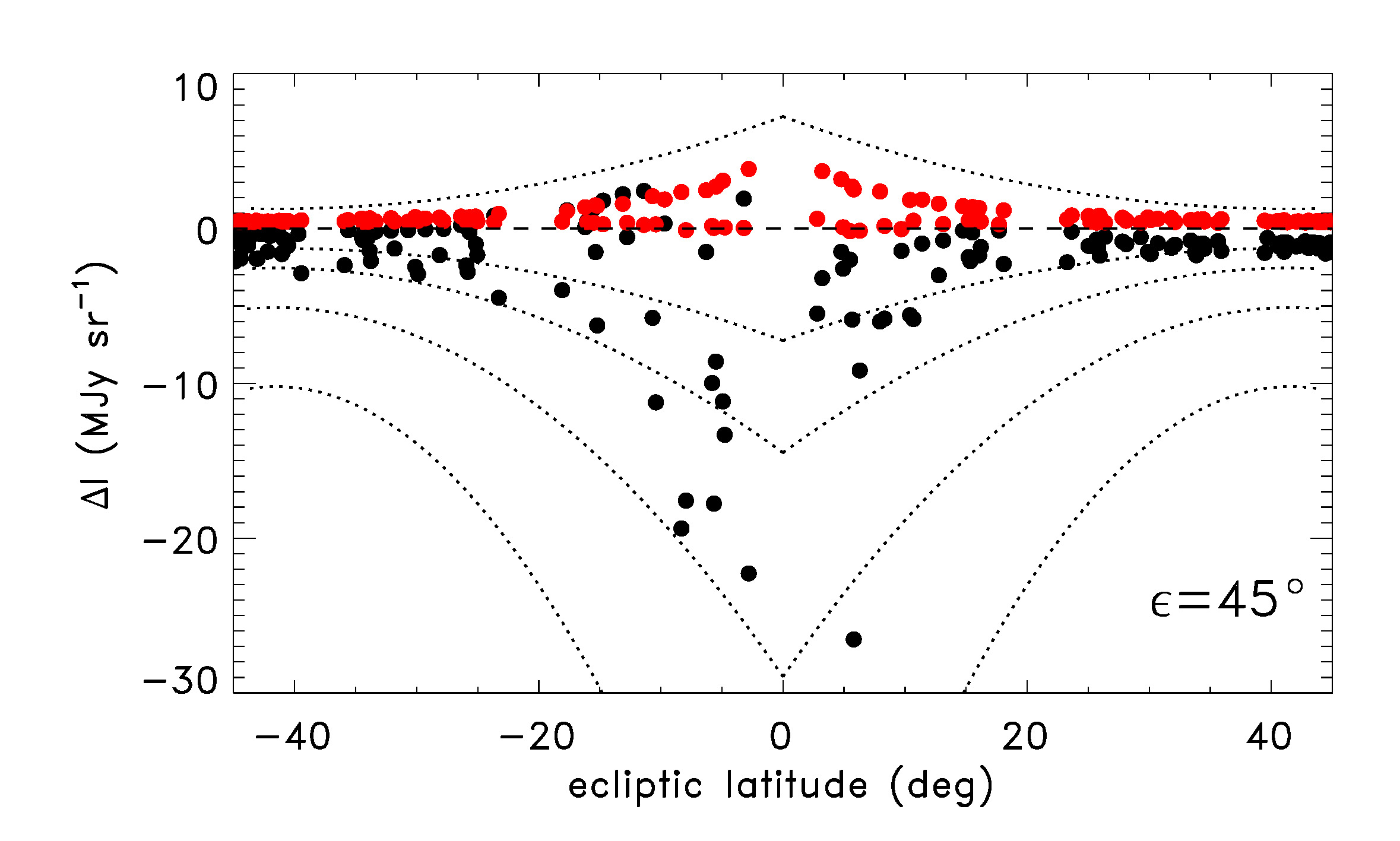}
    \caption{Difference in surface brightness predicted by the IPAC V1 and V4 models (black points) and the IPAC V1 and ZodiPy models (red points), at $\epsilon$\,=\,30\degr\ and 45\degr\ solar elongations.
    Dotted curves mark the 10, 20, 40 and 80\% differences.}
    \label{fig:sfbdiff}
\end{figure}

The comparison was performed for the first day of each month in 2026 using 12 positions distributed uniformly along a circle at a constant angular distance of $\epsilon = 30^\circ$ or 45\degr\ from the Sun, as viewed from Earth. 
A comparison of the predicted surface brightness values is presented in Fig.~\ref{fig:sfb} showing that, when the IPAC V1 model is adopted as the reference, the IPAC V4 predictions can differ substantially, by 10--80\%, with the largest discrepancies occurring at high surface brightness levels corresponding to low ecliptic latitudes (Fig.~\ref{fig:sfbdiff}). In absolute terms, the differences can reach $\sim$100\,\mjysr. In contrast, ZodiPy predictions remain within $\sim$5\% of the IPAC V1 values across the full surface-brightness range considered. The largest deviations occur at low ecliptic latitudes, where ZodiPy generally predicts slightly lower surface brightness values than IPAC V1.

\begin{figure}
    \centering
    \includegraphics[width=0.99\linewidth]{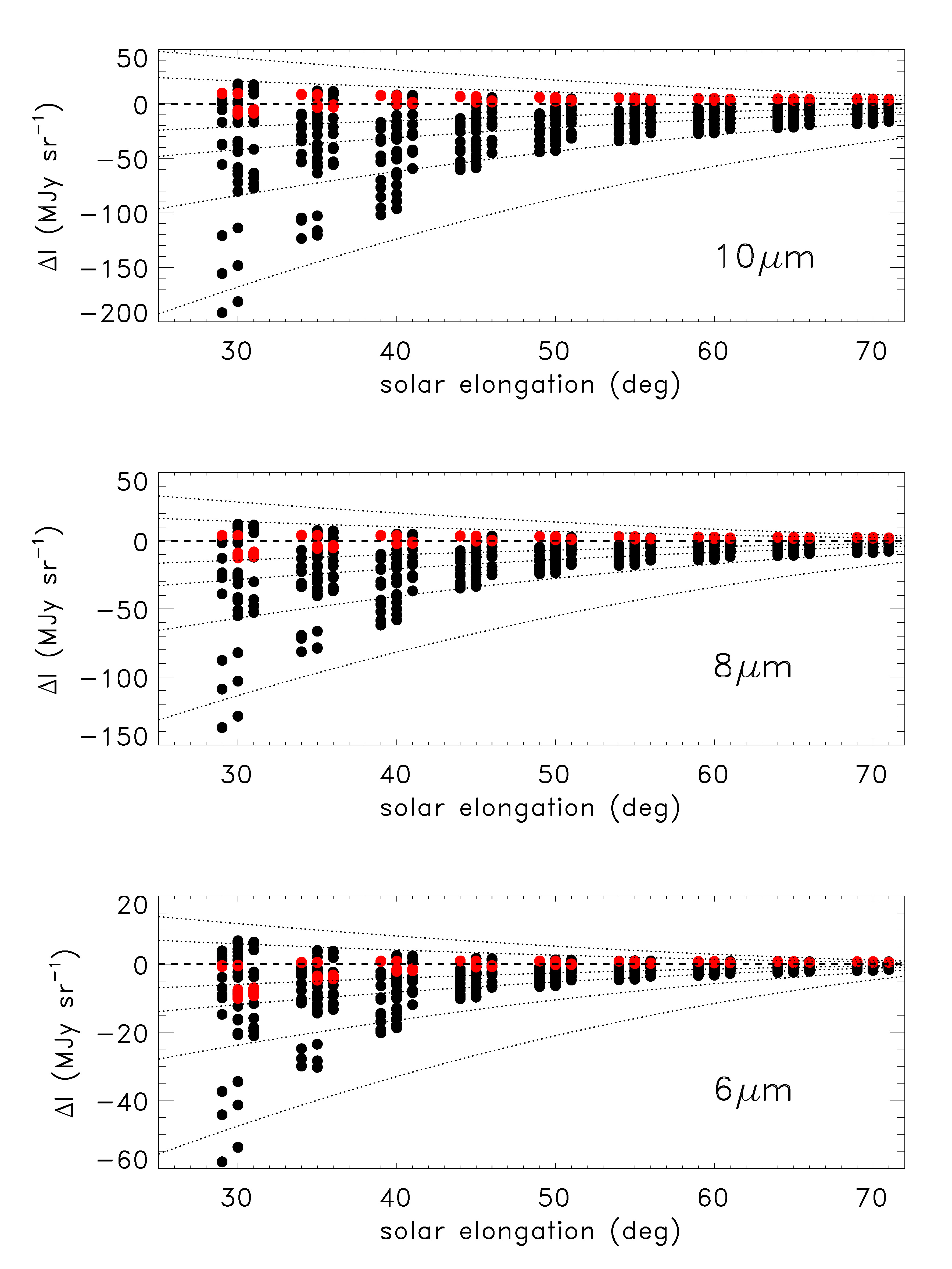}
    \caption{Difference of surface brightness predictions with respect to the IPAC V1 model in the Zodiacal could symmetry plane as a function of solar elongation, for 6, 8 and 10\,$\mu$m. Black and red dots represent the differences of the IPAC V4 and the ZodiPy models. Dotted curves mark the 10, 20, 40 and 80\% differences. }
    \label{fig:symmplane}
\end{figure}

We also compared the surface-brightness predictions of the different models along the zodiacal symmetry plane for solar elongations of $30^\circ \leq \epsilon \leq 70^\circ$ at wavelengths of 6, 8, and 10\,$\mu$m (Fig.~\ref{fig:symmplane}). The symmetry plane was defined following \citet{Kelsall1998}, with an inclination of $i = 2^\circ$ and an ascending node longitude of $\Omega = 78^\circ$. Relative to IPAC V1, the IPAC V4 predictions are up to $\sim$80\% higher, whereas ZodiPy agrees with IPAC V1 to within $\lesssim$10\% over the entire solar-elongation range considered. This behavior is consistent across all three wavelengths and closely mirrors the results obtained from the comparison at fixed solar elongation and varying ecliptic latitude.

{Regarding the $\lesssim10\%$ difference, ZodiPy implements the original \citet{Kelsall1998} DIRBE model and therefore shares the same underlying model as IPAC V1. The residual differences arise because the Kelsall model fitted independent spectral scale factors at the discrete DIRBE
wavelengths and thus does not uniquely specify predictions at the intermediate wavelengths considered here. ZodiPy and IPAC V1 use different prescriptions to interpolate these parameters, while IPAC V1 also adopts the \citet{Reach1997Icar..127..461R} treatment of the asteroidal dust bands. Their close agreement should therefore be interpreted as a consistency check between two implementations of the same underlying
model.}

\begin{figure}
    \centering
    \includegraphics[width=0.99\linewidth]{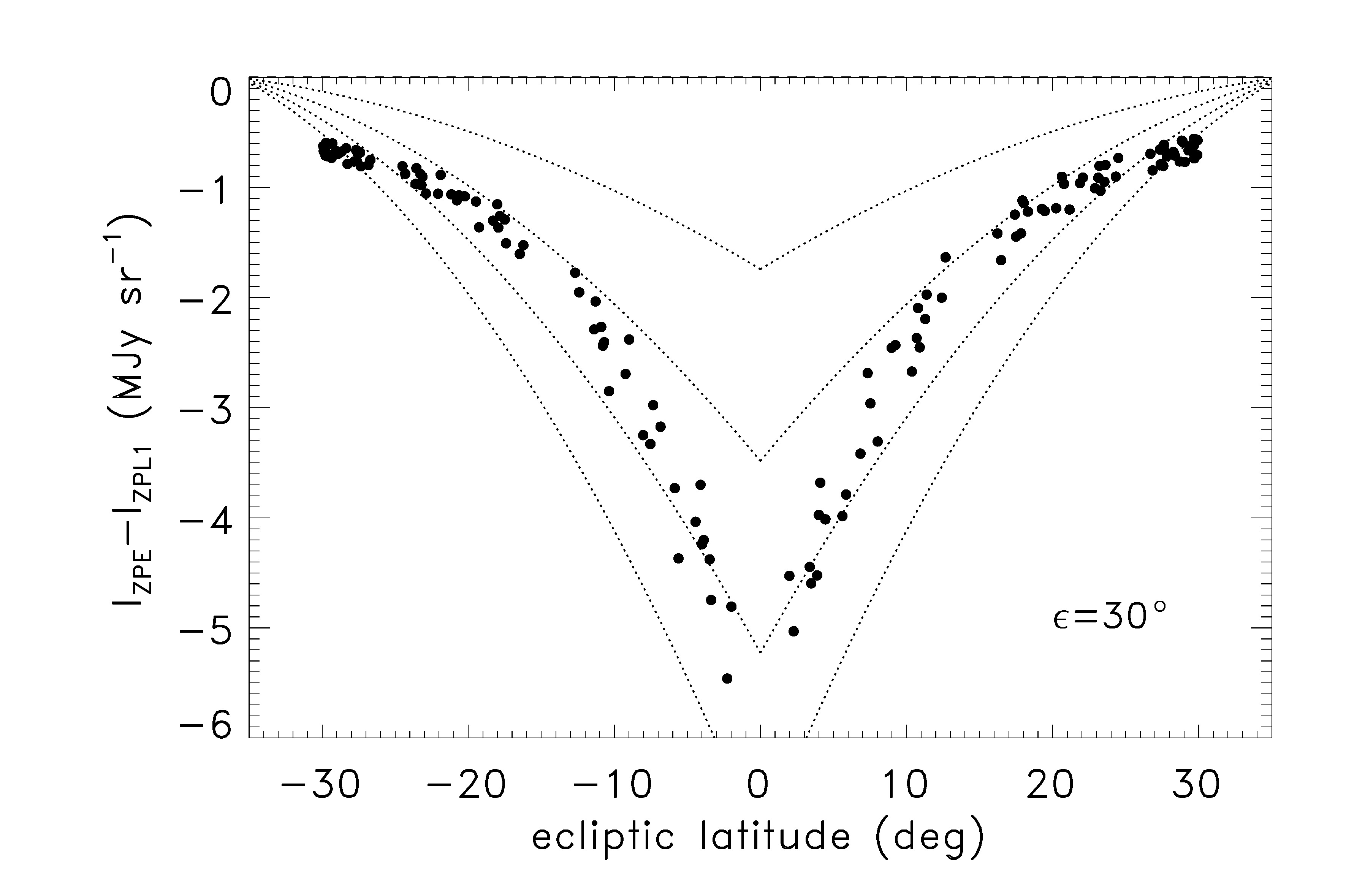}
    \caption{{Difference of surface brightness values predicted by ZodiPy at Earth and the Sun-Earth L1 point, for a set of pointing at $\epsilon$\,=\,30\degr\ solar elongation. Dotted curves mark the 1, 2, 3 and 4\% differences (from top to bottom). }
    }
    \label{fig:zpevsl1}
\end{figure}

As both NEO Surveyor and NEOMIR are planned to operate near the Sun--Earth L1 point, we also compared ZodiPy predictions obtained from Earth and from L1 for the same set of pointings at a solar elongation of 30\degr\ described above (Figure~\ref{fig:zpevsl1}). The predicted surface brightness at L1 is $\sim$2.7\% higher than at Earth. This ratio is essentially constant across the full range of ecliptic latitudes and surface-brightness values considered. The difference is consistent with the radial scaling of the zodiacal cloud adopted by the model \citep{Kelsall1998} and with the empirical surface-brightness dependence on heliocentric distance derived from the Helios observations (see Sect.~\ref{sect:helios} below).

\section{Validation against observational data \label{sect:zlmeas}}

Prior to COBE/DIRBE, a number of studies measured the zodiacal light at small solar elongations ($\epsilon \leq 60^\circ$). Although these measurements are limited in spatial coverage compared to the all-sky COBE/DIRBE observations, they provide a valuable opportunity to test the reliability of widely used zodiacal light models in a regime that remains otherwise weakly constrained. To our knowledge, however, no systematic comparison of these datasets with modern models has been carried out.
We identify three such datasets: visible-light measurements from the Helios probes, and mid-infrared observations from the Zodiacal Infrared Project (ZIP) rocket experiment and the Midcourse Space Experiment (MSX) satellite. 

In addition to the Helios, ZIP, and MSX measurements considered in this work, a number of other observations have probed the zodiacal light at small solar elongations. Using star-tracker images obtained by the \emph{Clementine} spacecraft, \citet{Hahn2002} constructed a visible-light map of the inner zodiacal cloud covering solar elongations of approximately $3^\circ$--$30^\circ$. Their analysis showed that the surface brightness increases steeply toward the Sun and implied a dust cross-sectional density varying approximately as $r^{-1.45}$. The Clementine observations were obtained with a navigation (star tracker) camera operating over a broad optical band spanning $\sim$0.4--1.1\,$\mu$m, providing high sensitivity to the diffuse zodiacal light. However, the broad, unfiltered spectral response complicates direct comparison with monochromatic model predictions and observations obtained in narrower photometric bands.

Similarly, observations obtained with the SOHO/LASCO coronagraphs and heliospheric imagers such as STEREO/HI and PSP/WISPR have provided extensive measurements of the F-corona and the inner zodiacal cloud at very small solar elongations. These observations have primarily been used to investigate the morphology, symmetry plane, temporal variability, and dust distribution of the near-Sun zodiacal cloud rather than to provide independent absolute surface-brightness measurements for validating zodiacal-light models \citep[e.g.,][]{Lamy2022}. Consequently, they are less suitable for quantitative comparisons with present-day zodiacal-light models than the Helios, ZIP, and MSX datasets analyzed here. Most recently
\citet{Tsumura2026} used Artemis II eclipse observations to study the optical F-corona, confirming its flattened morphology and finding that the radial dust distribution near the Sun is consistent with a power-law density profile of $n(r)\propto r^{-1.3}$, while highlighting discrepancies between observations and the ZodiSURF model at small solar elongations.

Below, we perform a comprehensive comparison of Helios, ZIP, and MSX observations with the DIRBE-based zodiacal light models IPAC V4 (based on the Wright/Gorjian formulation) and IPAC V1 (Kelsall et al./Price et al.). As shown in Sect.~\ref{sect:zlcomp}, the ZodiPy model \citep{San2024} agrees with IPAC V1 to within $\lesssim 5\%$ in most cases, which is smaller than the absolute calibration uncertainty of any of the datasets considered here. We therefore do not perform a separate evaluation of ZodiPy predictions. 

\subsection{Helios Space Probes \label{sect:helios}}

The Helios probes (Helios 1 and Helios 2, or Helios A and B) were joint NASA–DFVLR\footnote{Deutsche Forschungs- und Versuchsanstalt für Luft- und Raumfahrt} missions launched in 1974 and 1976 to explore the inner heliosphere from heliocentric orbits reaching perihelia as small as $\sim$0.29~au, well inside Earth’s orbit. The spin-stabilized spacecrafts carried a comprehensive in-situ payload to measure solar-wind plasma, magnetic fields, energetic particles, plasma waves, micrometeoroids, and zodiacal light, enabling the first sustained observations of solar-wind structure and variability close to the Sun. 

The zodiacal-light photometers aboard the Helios probes were designed to measure the brightness and color of scattered sunlight from interplanetary dust at heliocentric distances as small as $\sim$0.3~au. As described in the instrument reference papers \citep{1975RF.....19..264L,1976LNP....48...19P}, the photometer system consisted of three broadband channels with effective wavelengths near 370, 530, and 850\,nm, spanning the near-UV, visible, and near-infrared. Mounted on the spinning spacecraft, the photometers scanned great circles on the sky, with careful baffling and absolute radiometric calibration to suppress stray light and background contamination. The results of the Helios zodiacal emission measurements are summarized by \citet{1981A&A...103..177L}. 

While these measurements have not been performed in the infrared regime where NEO Surveyor and NEOMIR will operate, their comparison with the models at these wavelength still provides a type of consistency check of the models. As described in Sect.~\ref{sect:zlmodels}, the scattering and thermal emission components ($\sim$visible and infrared wavelengths) in the zodiacal light models are interlinked. Therefore the reliability of the scattered light component estimates at small solar elongations is also a measure of the reliability of the thermal emission component. 

In the light of the NEOMIR mission, here we specifically focus on Helios zodiacal light measurements that were performed at low solar elongations, typically within NEOMIR's solar elongation limits of 30\degr$\lesssim$\,$\epsilon$\,$\lesssim$\,70\degr. 
The Helios probes measured the zodiacal surface brightness regularly at solar elongations of $\epsilon$\,$\approx$\,30$^\circ$, and in some cases as small as $\epsilon$\,$\approx$\,15$^\circ$. 

\begin{figure}[ht!]
    \centering
    \includegraphics[width=0.99\linewidth]{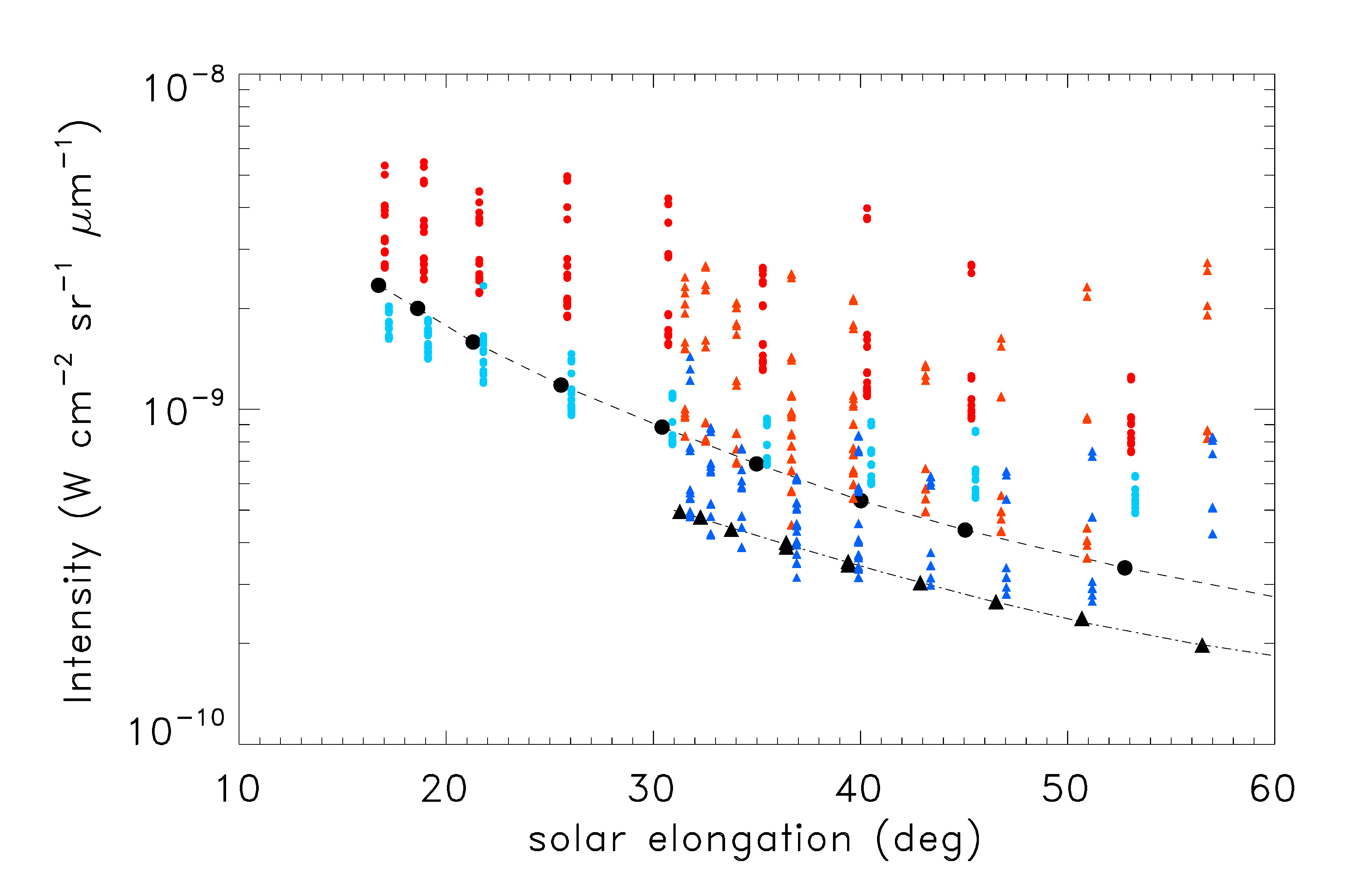}
    \caption{Surface brightness at 529\,nm, as observed by Helios 1 and 2 at aphelion ($\sim$1.0\,au). 
    Black dots (connected with a dashed curve) and triangles (dash-dotted curve) correspond to measurements at $|\beta|$\,=\,16 and 31\degr, respectively, and the same type of symbols are used for the preditions. Red and blue symbols mark the IPAC V4 and V1 predictions, respectively, and represent predictions for five epochs 1974-1976, as detailed in the text. 
    Note that data prediction points are slightly shifted in elongation for better visibility. }
    \label{fig:heliosap}
\end{figure}

The background surface brightness values published in \citet{1981A&A...103..177L} are comprised of the measurements in December 1974 and December 1975 (Helios A), and in January 1976, July 1976, and January 1977 (Helios B), when the probes were at their aphelion ($\sim$1\,au), allowing direct comparison with zodiacal measurements and model predictions from Earth. The measurements are performed down to $\epsilon$\,$\approx$\,15$^\circ$ in some cases, at specific ecliptic latitudes of $|\beta|$\,=\,16.2$^\circ$, and 31$^\circ$. Although the background values are consistent across different epochs, the values presented here in Fig.~\ref{fig:heliosap} are not associated with any specific epoch, and do not sample the same part of the zodiacal cloud at all observations. We calculated the predicted background values at 529\,nm (as in \citet{1981A&A...103..177L}) for the aphelion dates (1974/345 and 1975/359 for Helios A; 1976/015, 1976/201 and 1977/020 for Helios B\footnote{ephemeris dates are available from the NASA Open Data Portal: \url{https://data.nasa.gov/}}), and for all combinations of ecliptic coordinates of $\lambda\,=\,\lambda_\odot \pm \cos^{-1}(\cos{\epsilon}/\cos\beta)$ and $\pm|\beta|$. {Thus, the spread of the model points in Fig.~\ref{fig:heliosap} (and also in Fig.~\ref{fig:heliosph} below) also reflects geometrical ambiguity. }

The IPAC~V1 model predictions show a relatively low scatter throughout the year, with somewhat different values of the four positions per epochs corresponding to the solar elongation -- ecliptic latitude combinations. This scatter in the IPAC~V4 predictions is notably larger. The IPAC V4 predictions are significantly, 1.5-1.8-times higher than the Helios measurement levels at both the ecliptic latitudes ($|\beta|$\,=\,16.2$^\circ$ and 31$^\circ$).
The IPAC V1 predictions typically underestimate the Helios values; these differences are larger at very small (20 and 25$^\circ$) solar elongations, but are small (80-90\% of the observed value) for $\epsilon$\,=\,30--35$^\circ$. 
Overall, the IPAC V1 predictions (i.e. the Kelsall et al. model) show a good performance in predicting the zodiacal background in the solar elongation range of $\epsilon$\,$\gtrsim$\,30$^\circ$ relevant for NEOMIR.

\begin{figure}[ht!]
    \centering
    \includegraphics[width=0.99\linewidth]{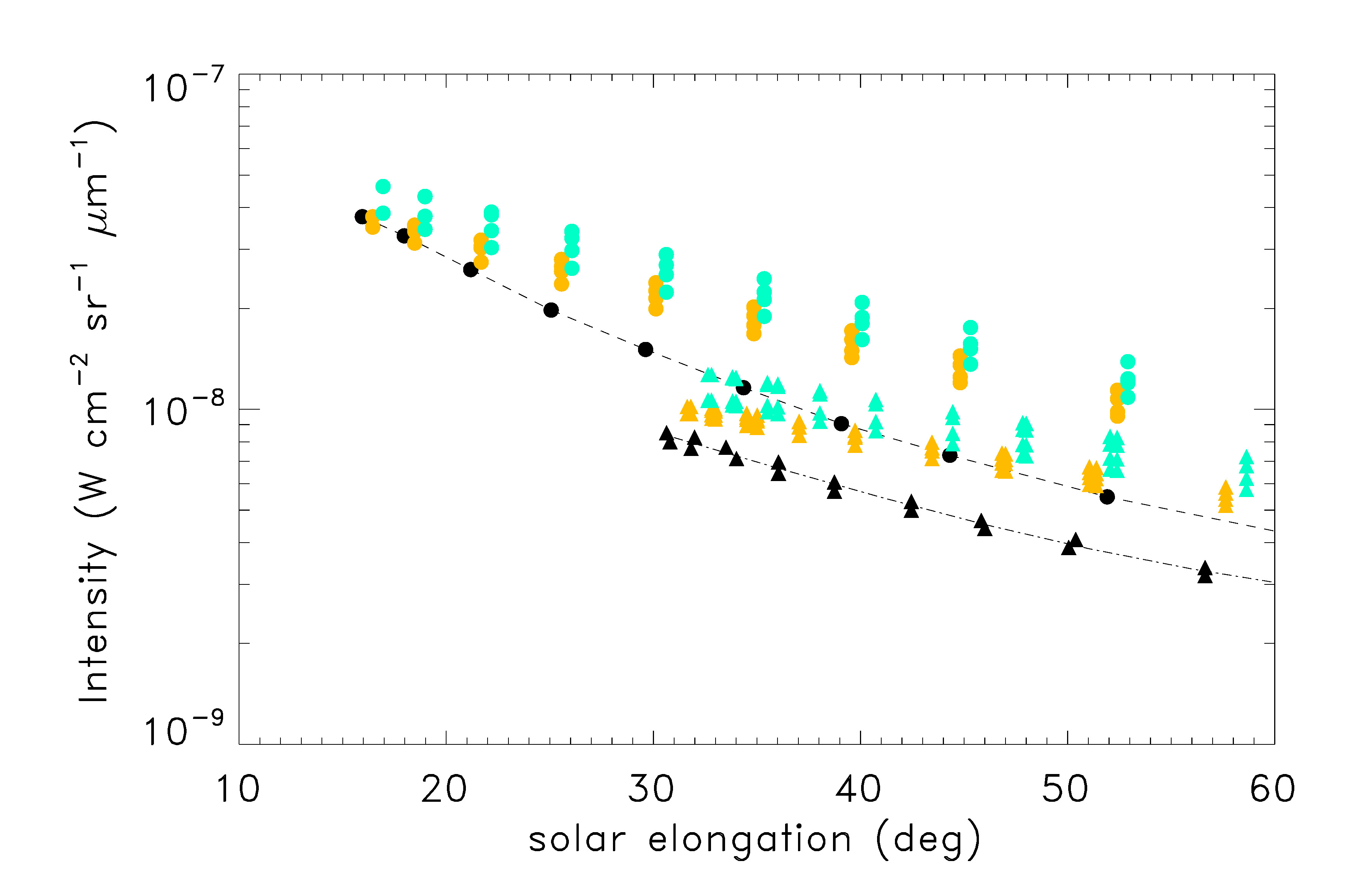}
\caption{Zodiacal-light surface brightness at 529~nm measured by Helios~1 and Helios~2 near perihelion ($r_{\rm h}\simeq0.3$~au), together with predictions calculated by ZodiPy for the position of each spacecraft. Orange and green denote Helios~1 and Helios~2, respectively, for both the measurements and their corresponding predictions. Circles and triangles denote $|\beta|=16^\circ$ and $31^\circ$, respectively. As in Fg.~\ref{fig:heliosap},  multiple predictions at a given elongation represent the possible viewing geometries consistent with the published Helios data, not observational scatter. IPAC V1 and V4 predictions are not shown because those interfaces
cannot evaluate the zodiacal-light models for an observer at these Helios spacecraft positions. Points are shifted slightly in elongation
for visibility.}
    \label{fig:heliosph}
\end{figure}

While there is presently no spacecraft planned in the foreseeable future for which zodiacal light intensity predictions would be important at very small heliocentric distances, we also compared 529\,nm Helios measurements taken at perihelion ($\sim$0.3\,au) and the predictions for the same configurations by ZodiPy, considering the same sky geometries as at aphelion. Helios data were obtained at two epochs, 1975/074 (Helios~1) and 1976/108 (Helios~2), and we used the Helios orbit files to obtain the ecliptic reference frame coordinates of the spacecrafts and the sky coordinates of the Sun at the perihelion dates. The results are presented in Fig.~\ref{fig:heliosph} for a sequence of measurements at constant ecliptic latitudes of $|\beta|$\,=\,16 and 31\degr, and for different solar elongations. While the predictions are quite close to the measured values at the smallest elongations, there is a considerable deviation for larger elongations at this small heliocentric distance. Overall the ZodiPy predictions overestimate the measured values by $\sim$100\% for larger solar elongations. This may be due to incorrect scaling laws applied in the Kelsall/ZodiPy models. 
(Note that these predictions could only be calculated by ZodiPy for 0.3\,au, and not by the IPAC V1 and V4 models). 

\subsection{Midcourse Space Experiment (MSX) scans}

\begin{figure*}[ht!]
    \centering
    \includegraphics[width=0.24\linewidth]{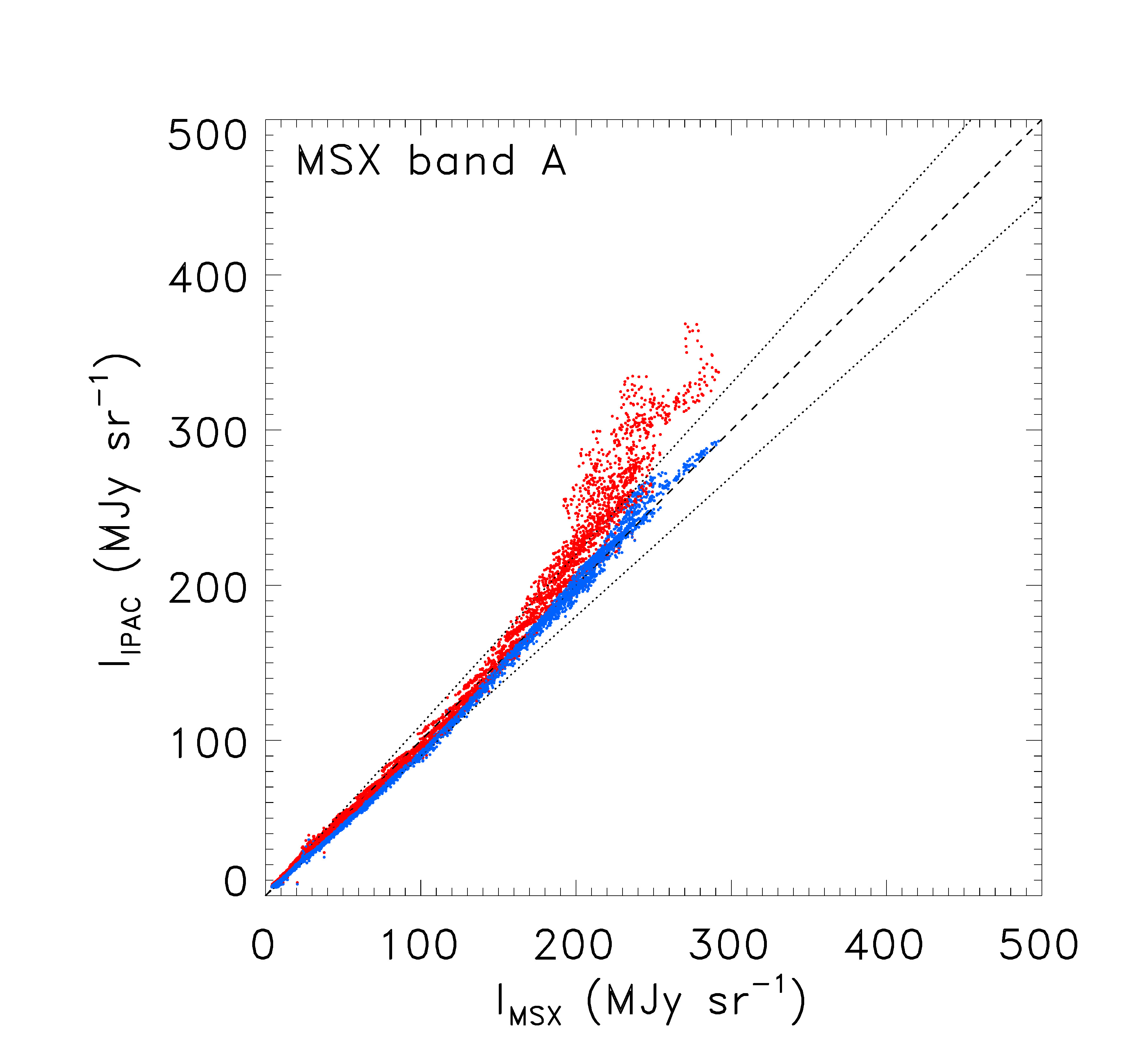} 
    \includegraphics[width=0.24\linewidth]{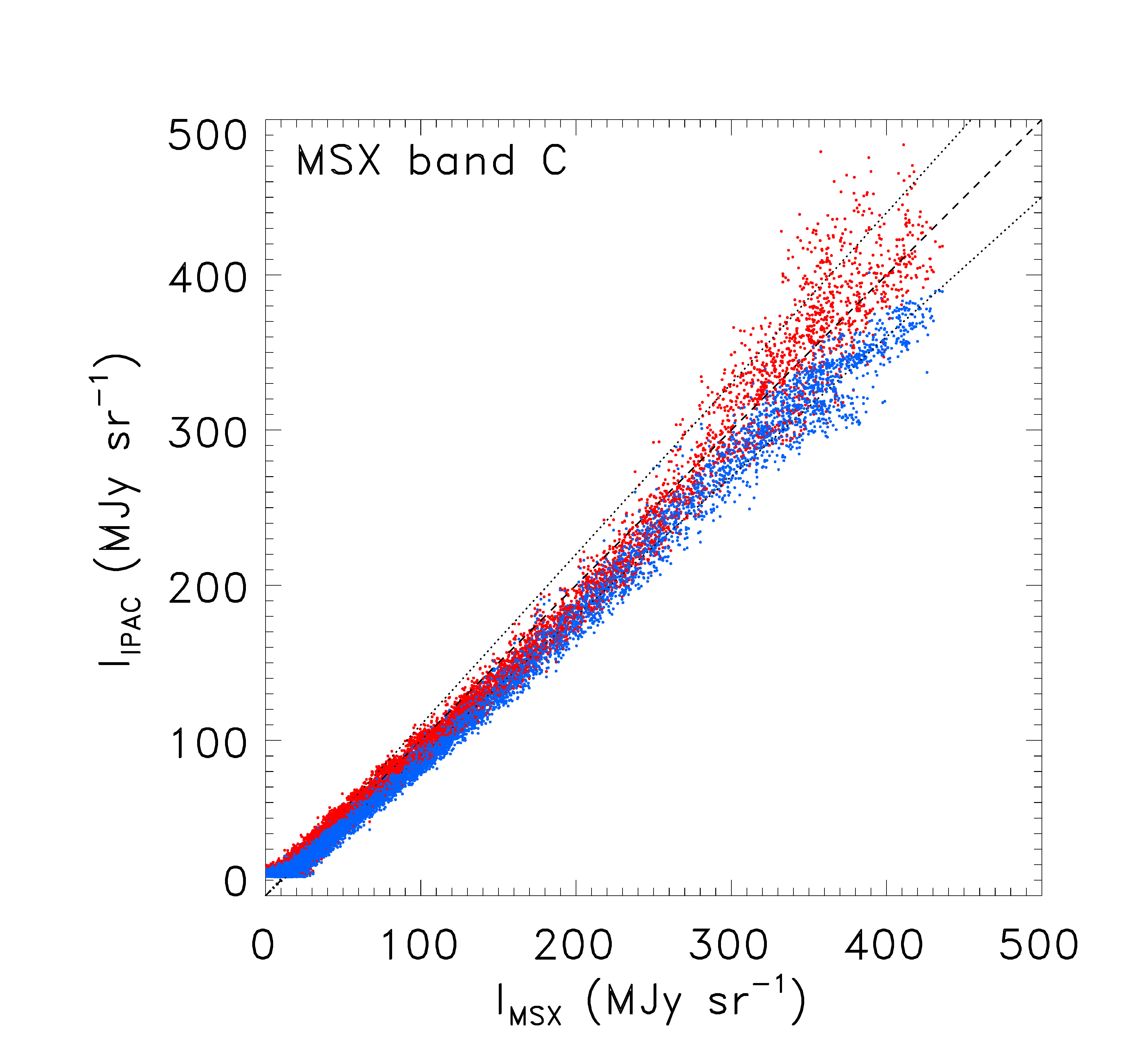} 
    \includegraphics[width=0.24\linewidth]{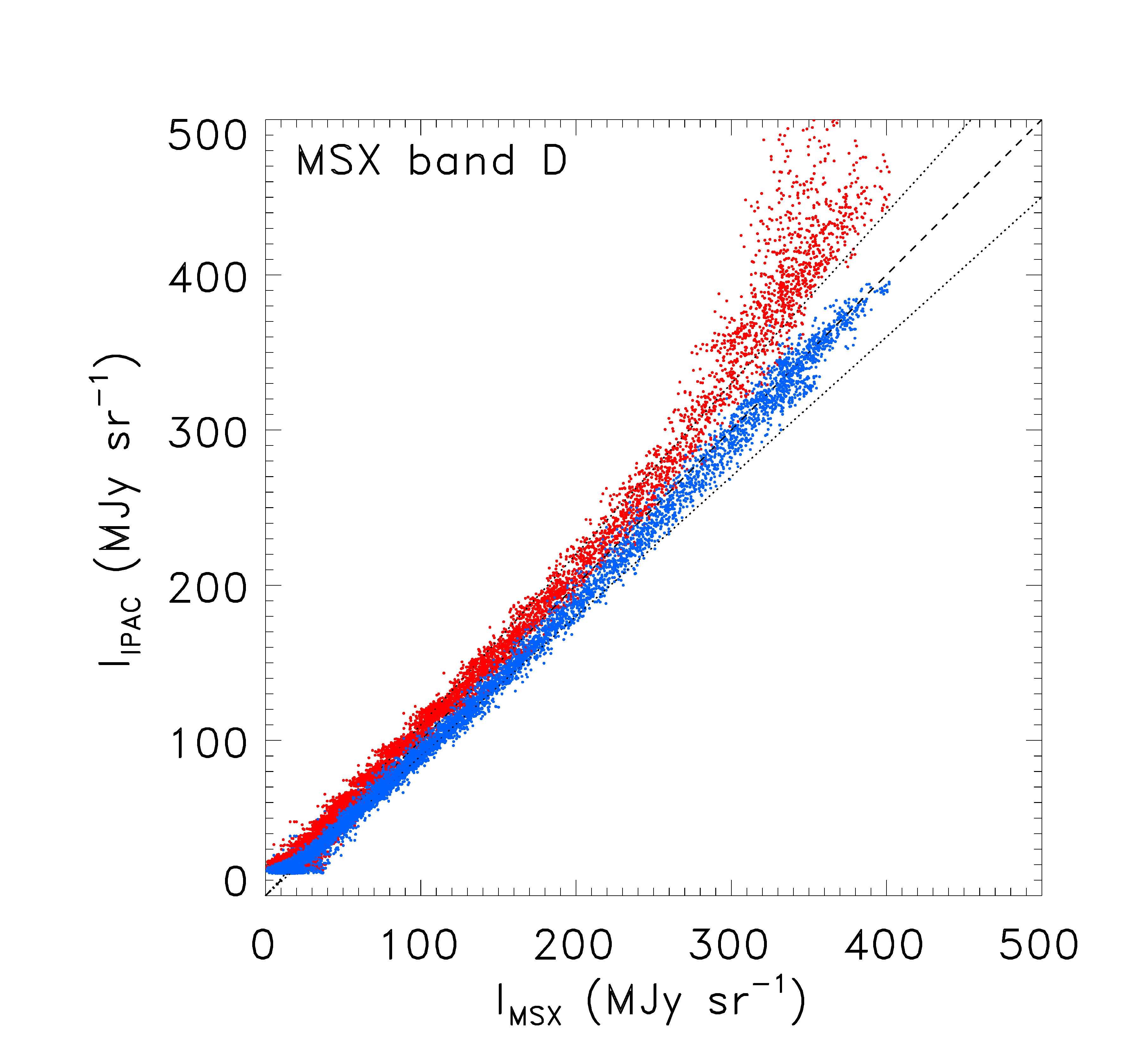} 
    \includegraphics[width=0.24\linewidth]{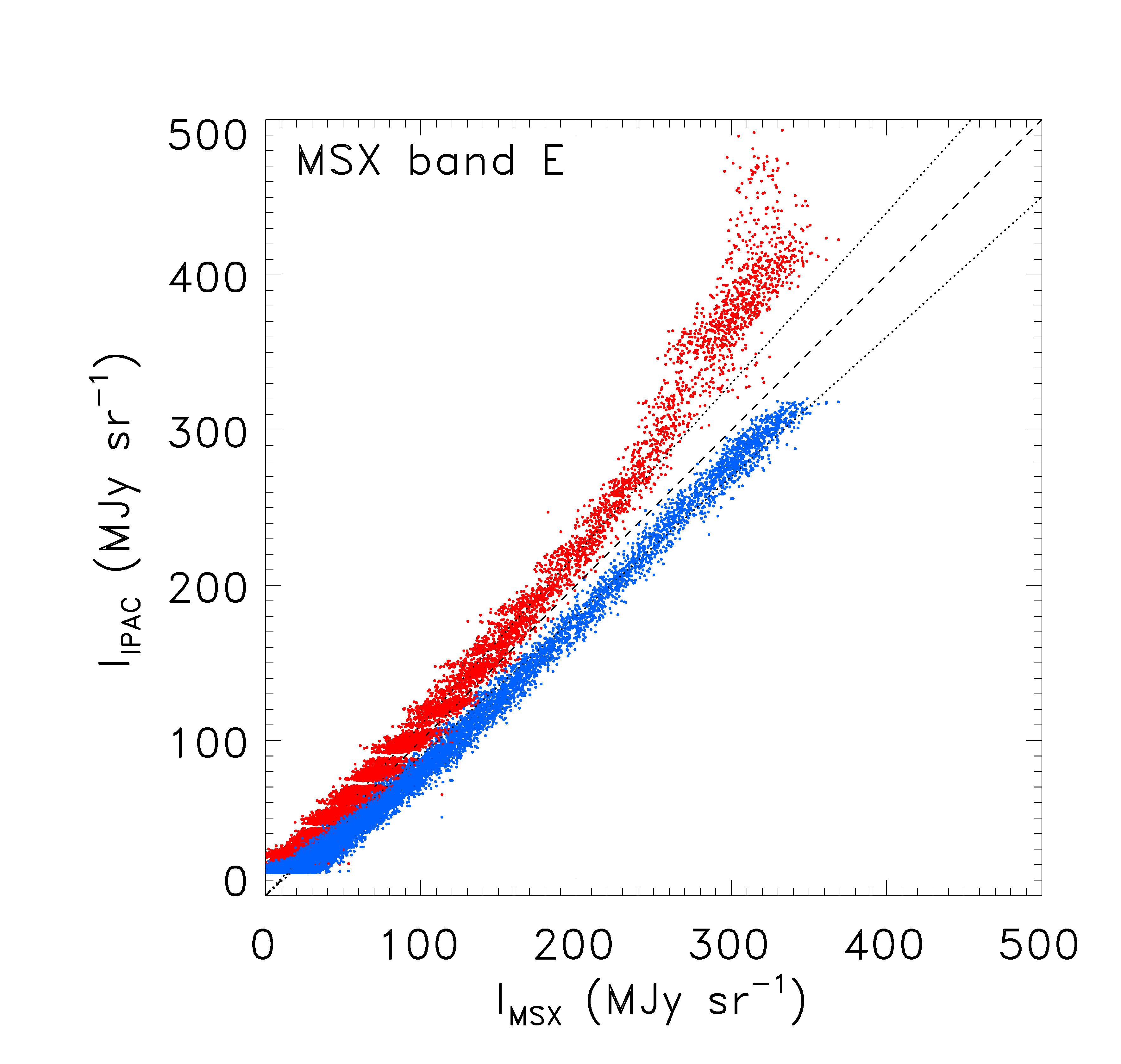}
    \caption{Comparison of MSX measurements and Wright / IPAC V4 (red) and Kelsall / IPAC V1 (blue) predictions for the MSX scans, in the A, C, D and E bands. The dashed line is the 1:1 ratio, dotted lines represent the $\pm$10\% deviations. }
    \label{fig:msxscatterplots}
\end{figure*}

The Midcourse Space Experiment (MSX) provided high-sensitivity, high–spatial-resolution measurements of the mid-infrared thermal emission from the zodiacal dust cloud during a nine-month observing campaign in 1996 \citep{Price2003}. Observations were carried out with the SPIRIT III cryogenically cooled infrared telescope, which employed discrete spectral bands centered at 8.3, 4.3, 12.13, 14.65, and 21.34\,$\mu$m (A, B, C, D and E bands) and was optimized for absolute measurements of diffuse emission.

The observing strategy extended mid-infrared coverage into regions inaccessible to earlier surveys (COBE/DIRBE and IRAS) because of solar exclusion constraints. While most measurements were performed at large solar elongations, the so-called CB01 scans consisted of five ecliptic pole-to-pole scans that came within 25–30$^\circ$ of the Sun. The data from these scans are available from the Planetary Science Institute (PSI) webpage\footnote{\url{https://sbn.psi.edu/pds/archive/msx.html}}.

We used the CB0103, CB0107, CB0108, CB0109, and CB110 scans in the A, C, D, and E bands, observed in late-July -- mid-August in 1996. Each scan typically contains 200-300 thousand individual pointings with corresponding in-band radiance values ($\mathrm{[W\,cm^{-2}\,sr^{-1}}$]). We used the data stored in \texttt{CORRECTED\_DATA\_VALUE}, which are corrected for Earth's scattered light. To avoid potential problems with interpolation, from each scan we directly selected 10\,000 pointings and retrieved the corresponding IPAC Background Model V1 and V4 predictions. A fraction of pointings was discarded due to unreliable subtraction of the Earth's scattered light contribution. This was effectively done by identifying pointings where \texttt{NRER\_VALUE}\,$\leq$0.1$\times$\texttt{DATA\_VALUE}, i.e. a low straylight contribution. Note that \citet{Price2003} claims 9, 8, 9 and 15\% accuracy in absolute radiance in the A, C, D and E bands, respectively. 

\begin{table}[ht!]
    \centering
    \begin{tabular}{cccccc}
    \hline
     Band & $\lambda_c$ & $\Delta\lambda$ & C$_{300}$ & C$_{350}$ & C$_{\Delta\lambda}$ \\
          & ($\mu$m) & ($\mu$m) & \multicolumn{3}{c}{[MJy\,sr$^{-1}$]/[W\,cm$^{-2}$\,sr$^{-1}$]} \\ \hline
      A &  8.28 & 3.36 & 6.97$\times10^{10}$ & 7.22$\times10^{10}$ & 7.13$\times10^{10}$ \\
      C & 12.13 & 1.72 & 2.87$\times10^{11}$ & 2.88$\times10^{11}$ & 2.86$\times10^{11}$ \\
      D & 14.65 & 2.23 & 3.22$\times10^{11}$ & 3.23$\times10^{11}$ & 3.11$\times10^{11}$ \\
      E & 21.34 & 6.24 & 2.50$\times10^{11}$ & 2.49$\times10^{11}$ & 2.48$\times10^{11}$ \\ \hline
    \end{tabular}
    \caption{Transformation coefficients for the MSX bands used in our study. C$_{300}$ and C$_{350}$ are calculated using convolutions of the filter transmission profiles and spectral energy distributions with a common black body temperature of 300\,K, and 350\,K. The C$_{\Delta\lambda}$ coefficients, obtained using the effective filter width and assuming a flat spectrum, are also shown for comparison. }
    \label{table:msxcoeff}
\end{table}

We applied the coefficients listed in Table~\ref{table:msxcoeff} to transform the band-integrated radiance (stored in units of [W\,cm$^{-2}$\,sr$^{-1}$]) to monochromatic surface brightness ([MJy\,sr$^{-1}$]), using the filter transmission curves available at the PSI website and assuming a blackbody spectral energy distribution with a temperature of 300\,K. This choice accounts for the smaller heliocentric distances sampled by these scans, which correspond to higher dust temperatures. 
A detailed description of the temperature structure of the zodiacal emission at larger solar elongations is discussed in \citet{Abraham1999ESASP.427..145A} and \citet{Leinert2002A&A...393.1073L}. 
{\citet{Kelsall1998} adopted a blackbody temperature of
$T_0=286\,{\rm K}$ at 1\,au and fitted the heliocentric-distance exponent, obtaining
$T(r_{\rm h})=T_0(1\,{\rm au}/r_{\rm h})^{0.467}$. }
Physically reasonable but different spectral energy distributions, for example adopting blackbody temperatures up to $\sim$350\,K to account for even higher temperatures along some lines of sight, lead to coefficients that differ only by $\lesssim$5\% in all bands. Scatter plots of the predicted and observed surface brightness values are presented in Fig.~\ref{fig:msxscatterplots}.

\begin{figure}[ht!]
    \centering
    \includegraphics[width=0.99\linewidth]{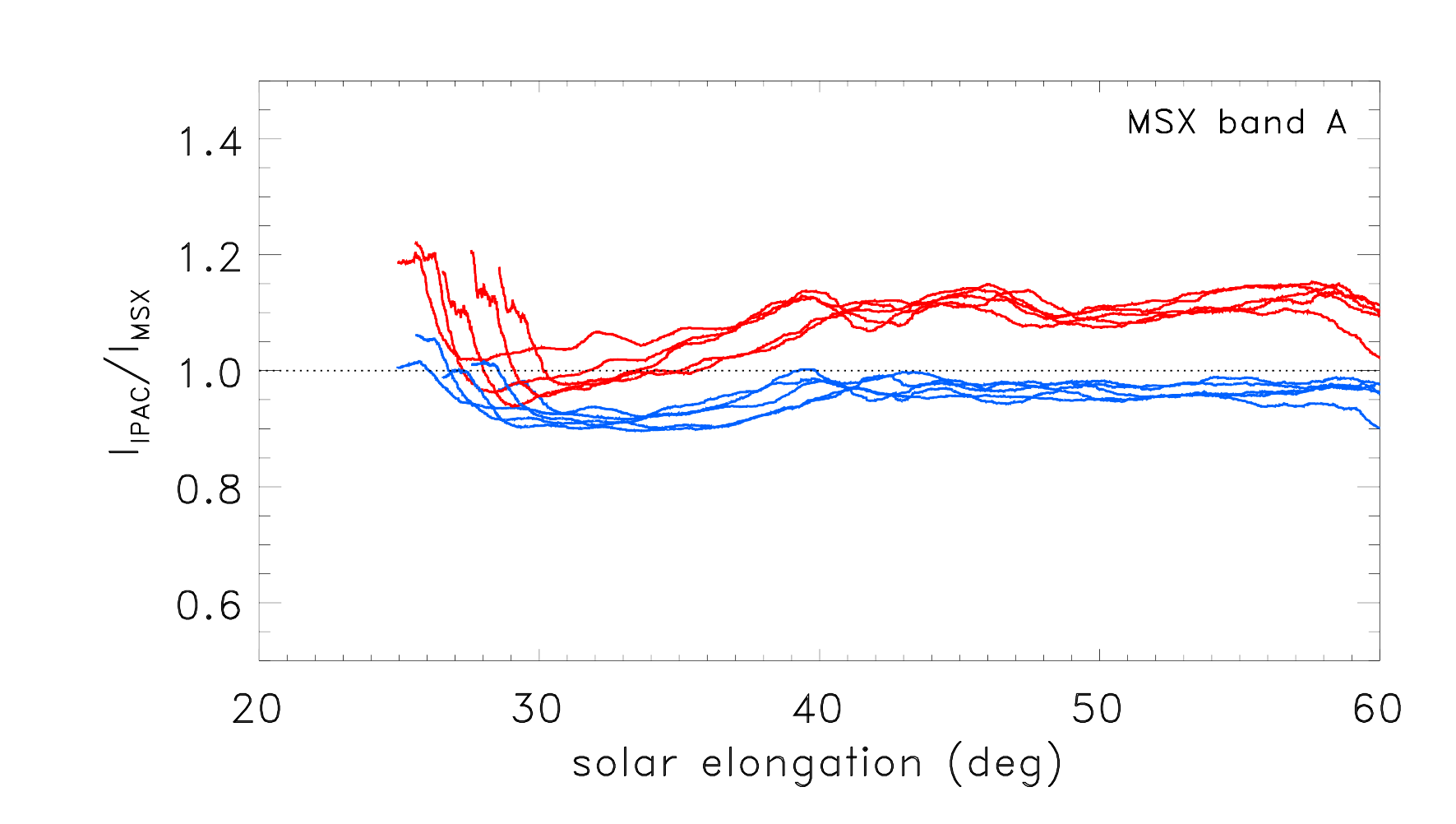}
    \includegraphics[width=0.99\linewidth]{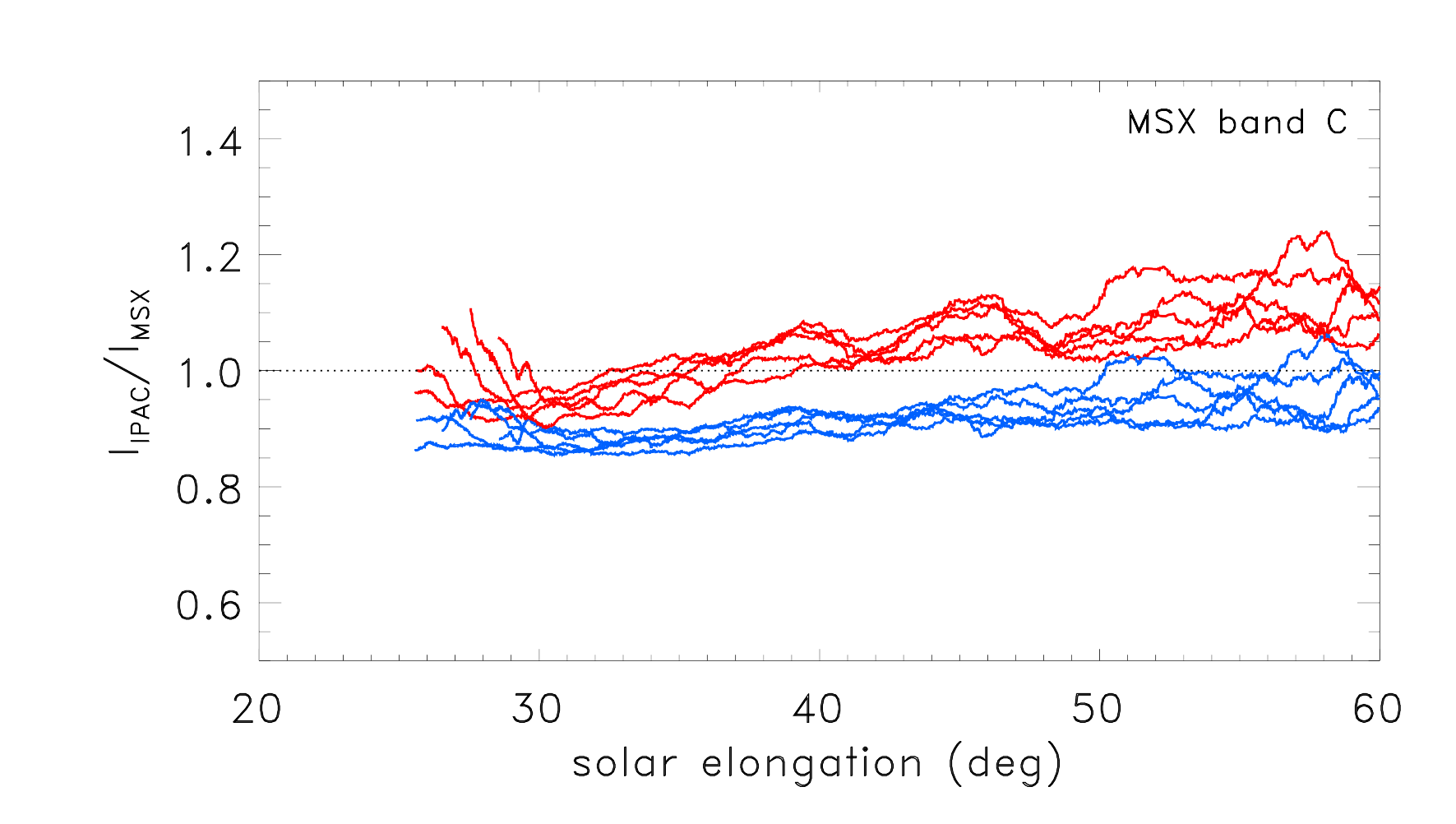}
    \includegraphics[width=0.99\linewidth]{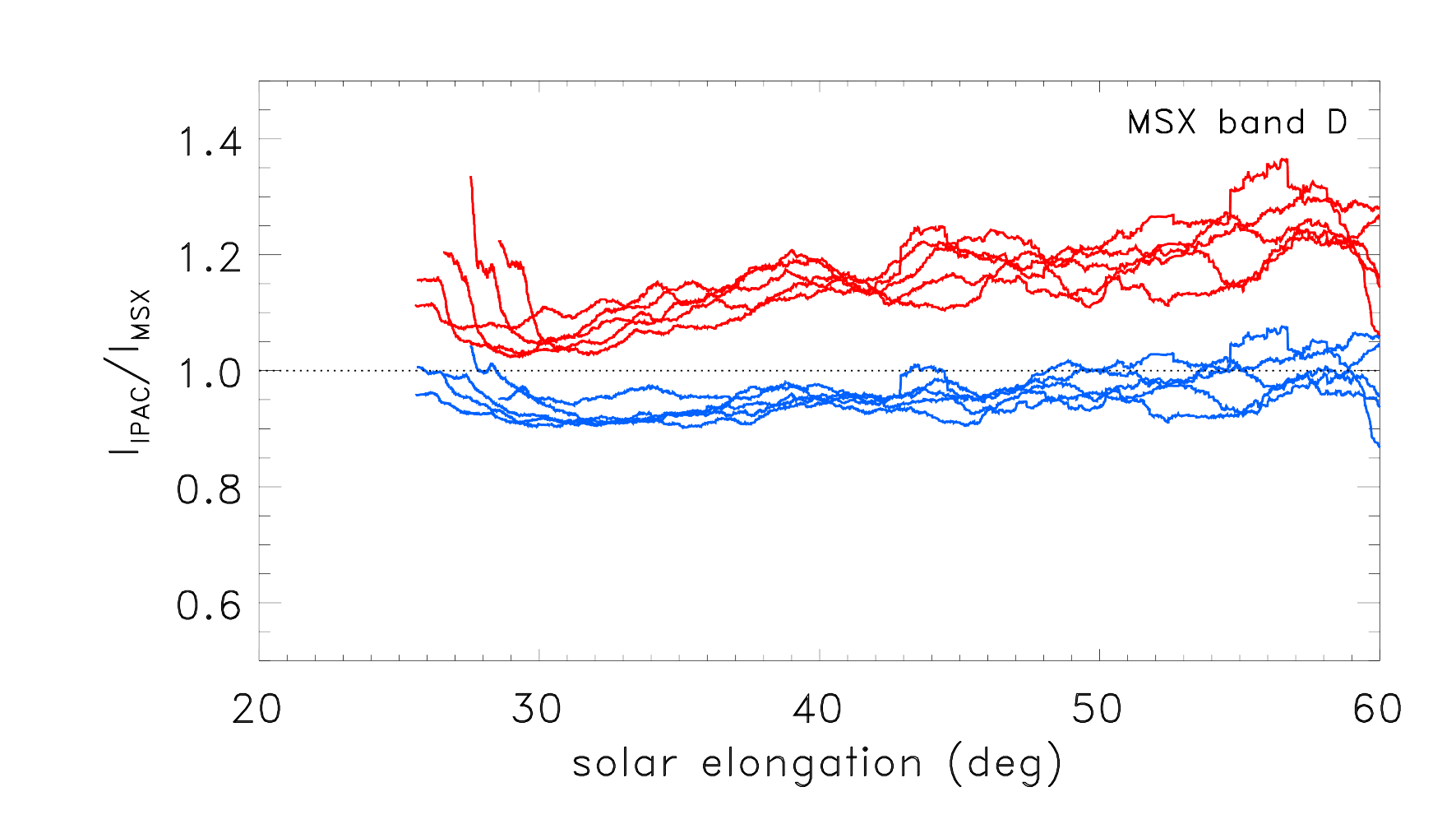}
    \includegraphics[width=0.99\linewidth]{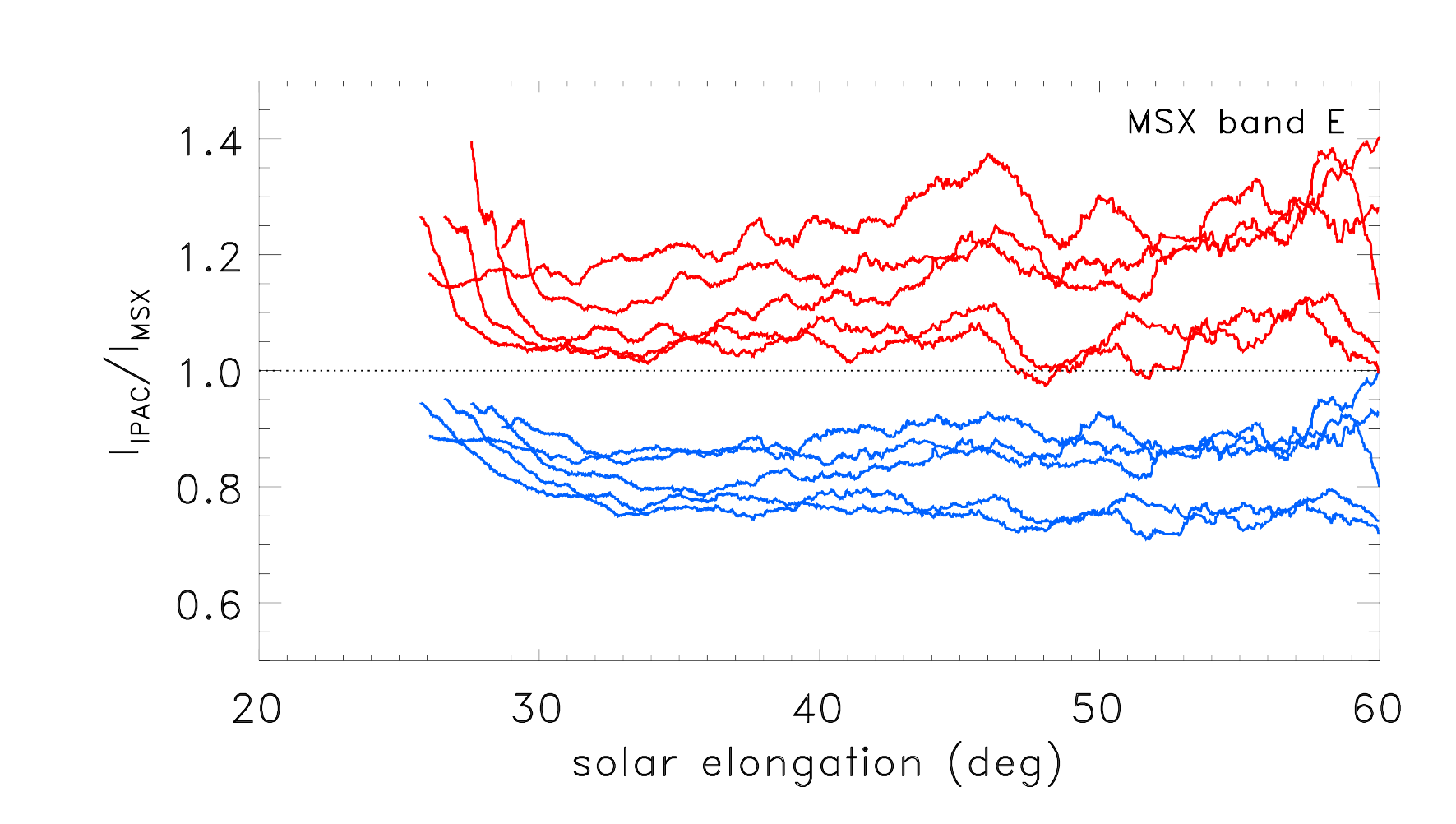}
    \caption{Median ratio of the IPAC predictions to the MSX measurements as a function of solar elongation. The curves of the five scans are shown separately. Red and blue curves correspond to IPAC V4 and V1, respectively. The panels correspond to the bands A, C, D and E, from top to bottom.}
    \label{fig:msxbands}
\end{figure}

For MSX scans, the 8.28\,$\mu$m (A) band is the most relevant for comparison with the wavelength ranges of NEOMIR and NEO Surveyor. The IPAC-to-MSX ratios exhibit similar behavior in all bands (Fig.~\ref{fig:msxbands}). In band A, IPAC V4 overestimates the surface brightness by approximately 20\% at $\epsilon \lesssim 30^\circ$, although individual pointings show larger deviations. The ratios approach unity near $\epsilon \approx 30^\circ$ and gradually increase to an overestimation of about 10\% at larger elongations. IPAC V1 generally provides a better match to the observations, with ratios remaining closer to unity over a broader range of solar elongations.

The remaining bands show the same overall trends, but with an increasing scatter towards longer wavelengths. Across all bands, IPAC V4 systematically predicts higher surface brightnesses than observed. In bands A and D, IPAC V1 provides a substantially better agreement with the data. In band E, IPAC V4 overestimates the surface brightness by approximately the same amount that IPAC V1 underestimates it over a wide range of elongations; however, at the smallest solar elongations, IPAC V1 remains close to the observations, whereas IPAC V4 significantly overpredicts the brightness.

\subsection{Zodiacal Infrared Project (ZIP)}

In the Zodiacal Infrared Project \citep[ZIP, see][]{Murdock1985AJ.....90..375M} the infrared zodiacal emission was measured at wavelengths between 2 and 30\,$\mu$m with a cryogenically cooled absolute radiometer from a rocket platform, covering solar elongation angles between 22$^\circ$\,$\leq$\,$\varepsilon$\,$\leq$\,180$^\circ$ and elevation angles between the North Ecliptic Pole and 60$^\circ$ south ecliptic latitude. The observing strategy relied on controlled scanning of the sky through changes in solar elongation and ecliptic latitude, achieved by either payload rotation or stepped pointing during flight, producing repeated cuts across the ecliptic plane. 
Most of the small solar elongation ($\epsilon$\,$\leq$45$^\circ$) measurements were performed during the flight on July 31, 1981. Surface brightness measurements as a function of solar elongation and ecliptic latitude are available at 10.9 and 20.9\,$\mu$m in \citet{Murdock1985AJ.....90..375M}, therefore we use these wavelengths to compare the ZIP results with the predictions. ZIP data are presented in in-band $\mathrm{W\,cm^2\,\mu m^{-1}\,sr^{-1}}$ surface brightness units, which can be directly transformed to $\mathrm{MJy\, sr^{-1}}$, or vice versa. ZIP data are calibrated using a reference spectral energy distribution of a 6000\,K black body. This results in a colour correction factor of $K\approx$\,0.9 for a 300\,K black body in the relatively broad 10.9\,$\mu$m filter (filter width of 4.38\,$\mu$m), and K values even closer to unity for higher dust temperatures, representative of small solar elongations. As the 20.9~$\mu$m filter is narrower (2.11~$\mu$m), the colour corrections are very close to unity. As the exact temperature or spectral energy distribution of the background in the line of sight is not known, we chose a uniform colour correction factor of $K\,=\,0.9$ for 10.9\,$\mu$m, and $K\,=\,1$ for 20.9\,$\mu$m, however, this choice does not affect our final results noticeably. The absolute radiometric calibration of the ZIP data was established using a laboratory blackbody reference. Although the detector response was reproducible to within approximately 5\%, the uncertainty in the transfer of the laboratory calibration to the absolute sky radiance, dominated by uncertainties in the reference blackbody emissivity and temperature, resulted in an estimated absolute calibration uncertainty of approximately 20\%. 

\begin{figure*}[ht!]
    \centering
    \hbox{
    \includegraphics[width=0.33\linewidth]
    {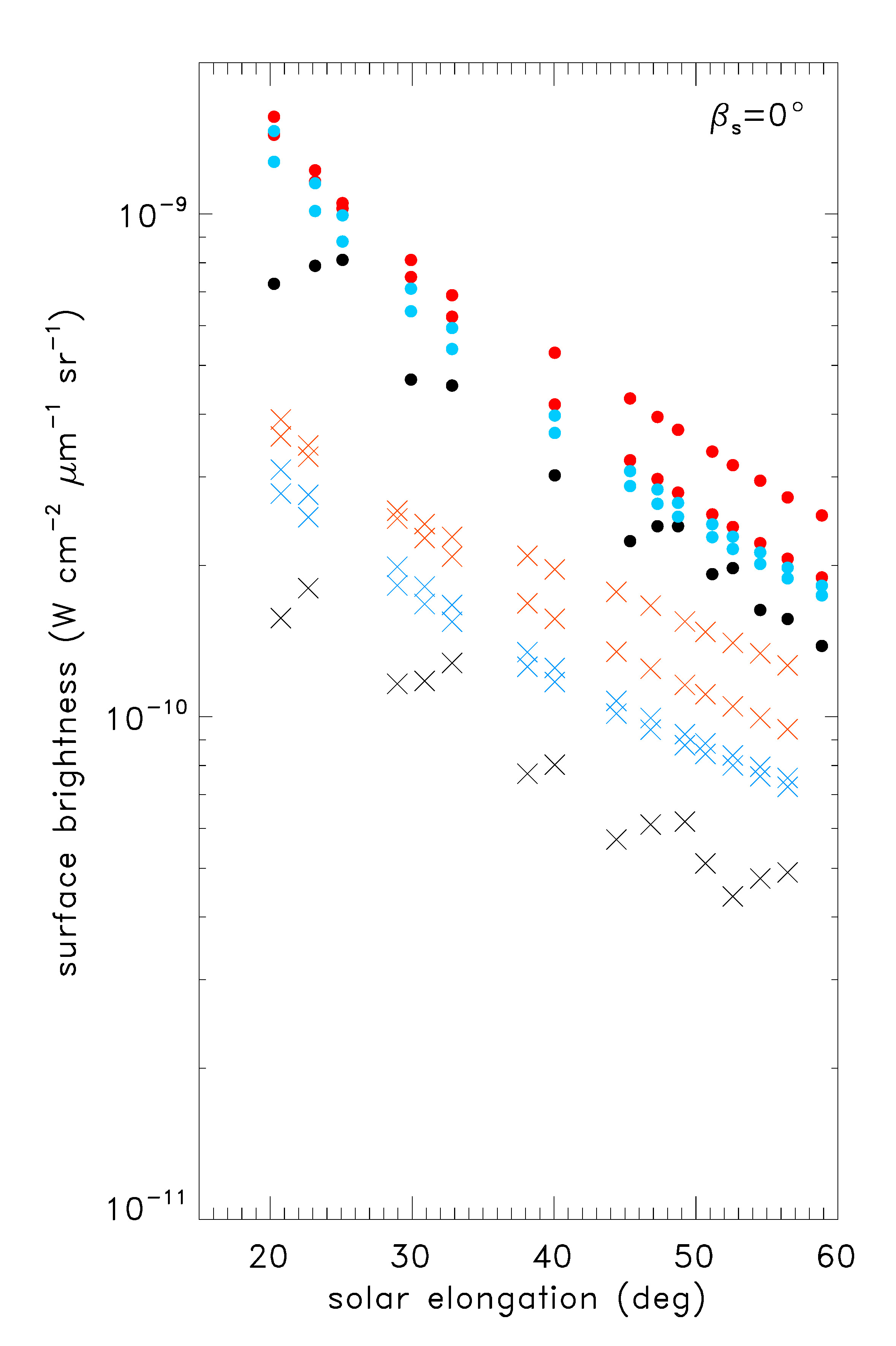}
        \includegraphics[width=0.33\linewidth]
    {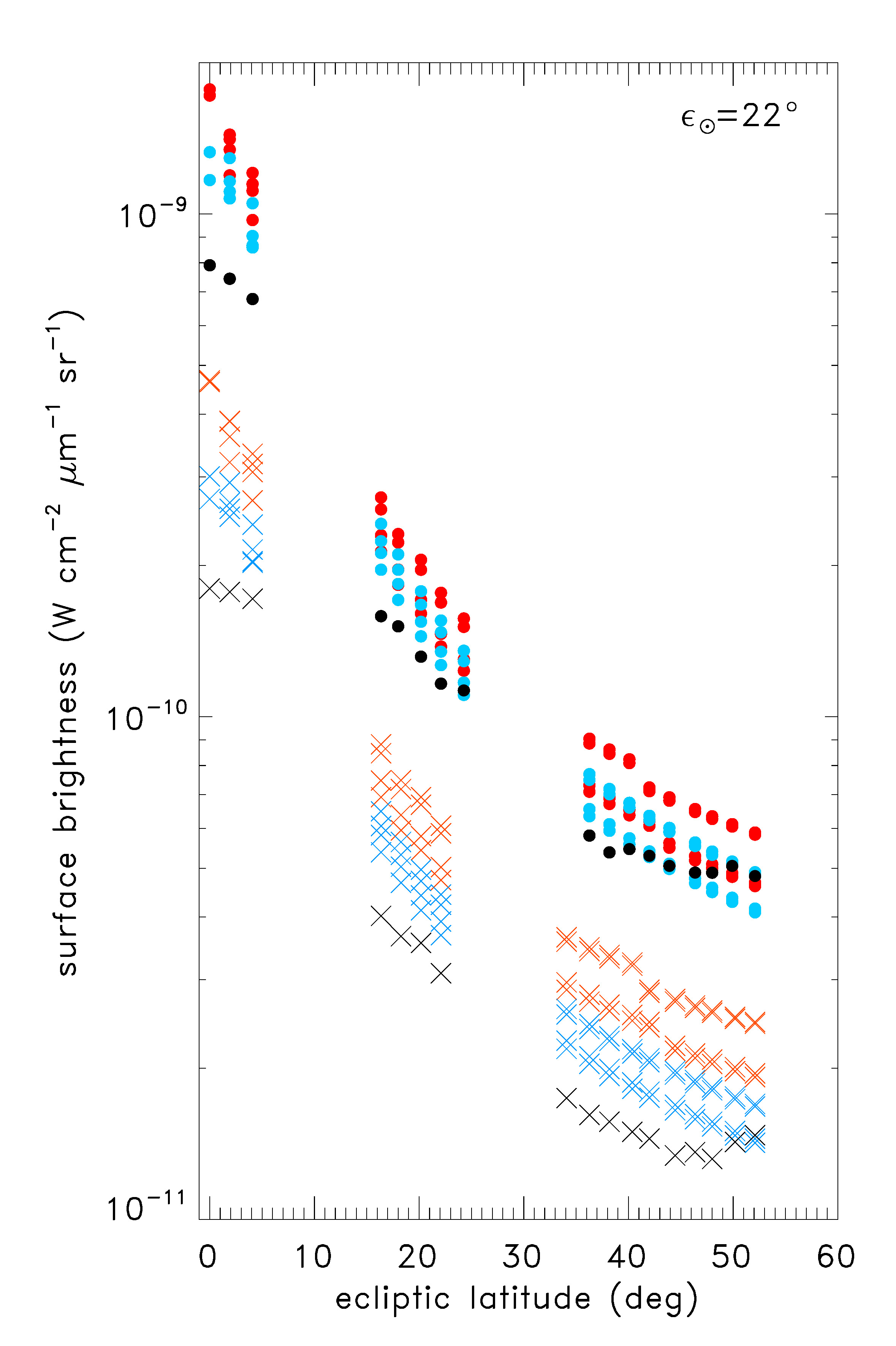}
        \includegraphics[width=0.33\linewidth]
    {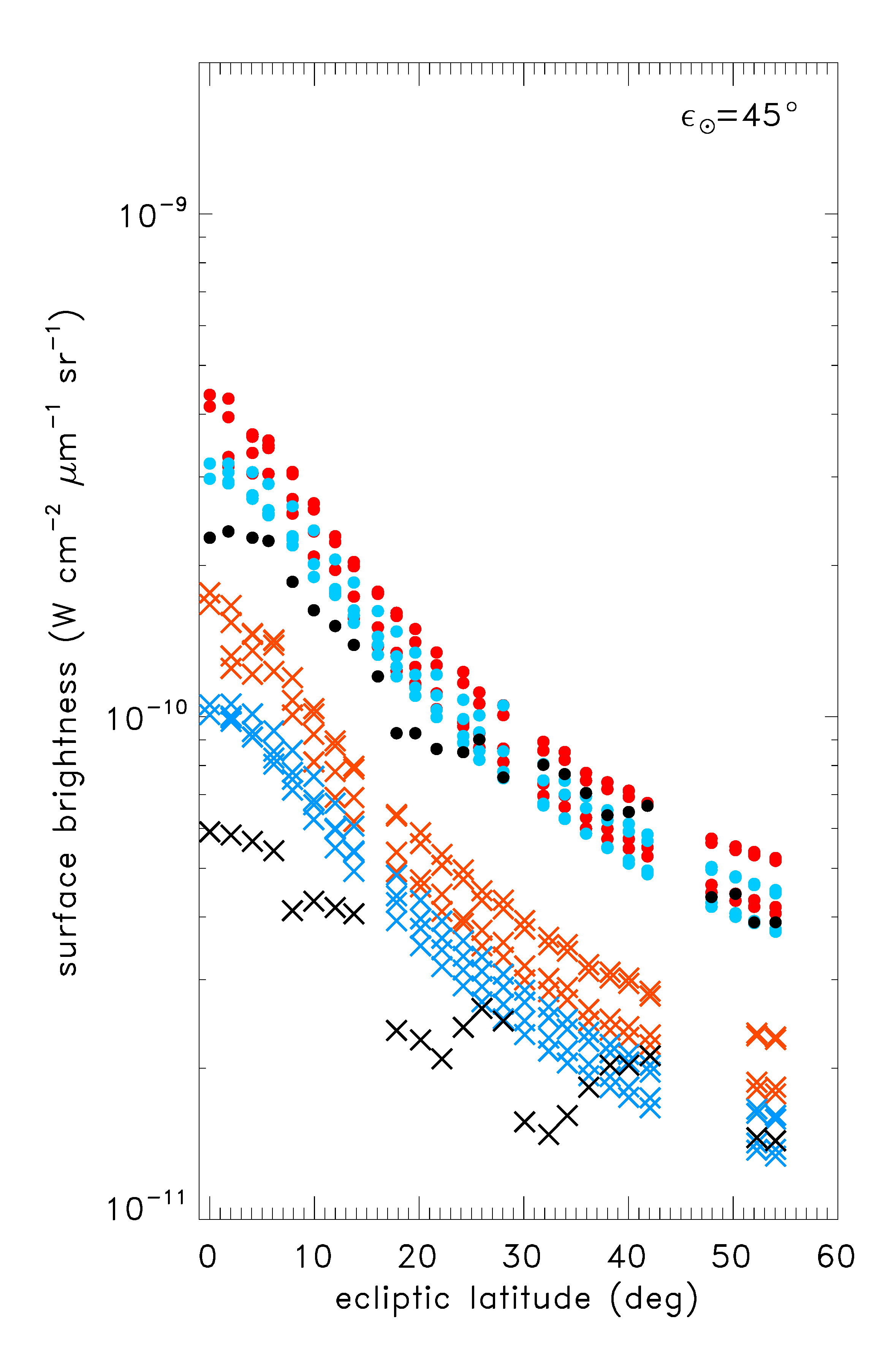}
    }
    \caption{Measured and model surface brightness of ZIP pointings. Dots and x-symbols mark the 10.9 and 20.9\,$\mu$m data, respectively. 
    Black, blue and red colours correspond to the measured, IPAC V1, and IPAC V4 models. Left panel: data in the zodiacal symmetry plane as a function of solar ecliptic elongation $\epsilon_\odot$\,=\,$|\lambda-\lambda_\odot|$. Middle panel: data at $|\lambda-\lambda_\odot|$\,=\,22$^\circ$, as a function of ecliptic latitude. Right panel: data at $|\lambda-\lambda_\odot|$\,=\,45$^\circ$, as a function of ecliptic latitude.}
    \label{fig:zipsfb}
\end{figure*}

\begin{figure*}[ht!]
    \centering
    \hbox{
    \includegraphics[width=0.33\linewidth]{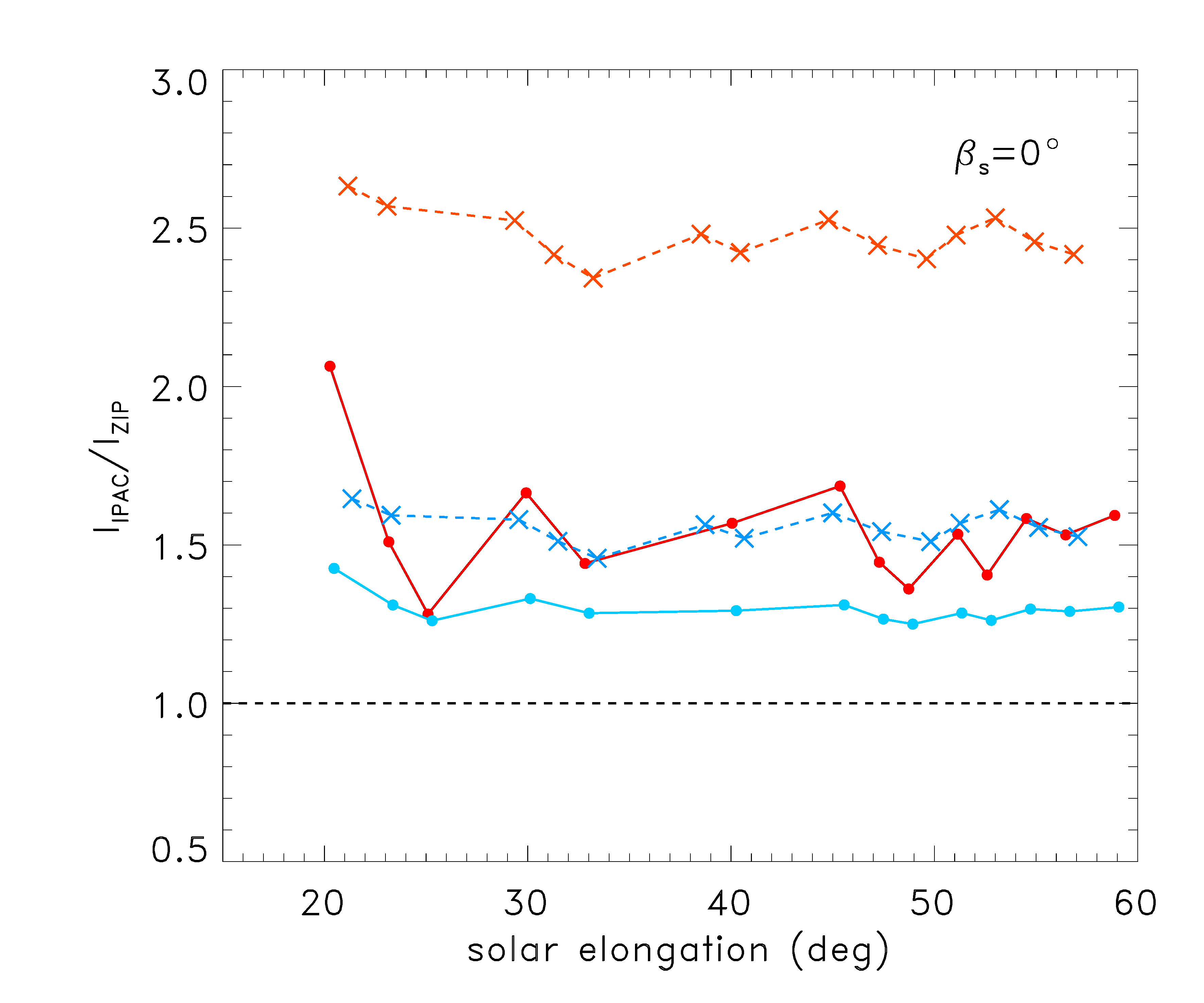}
    \includegraphics[width=0.33\linewidth]{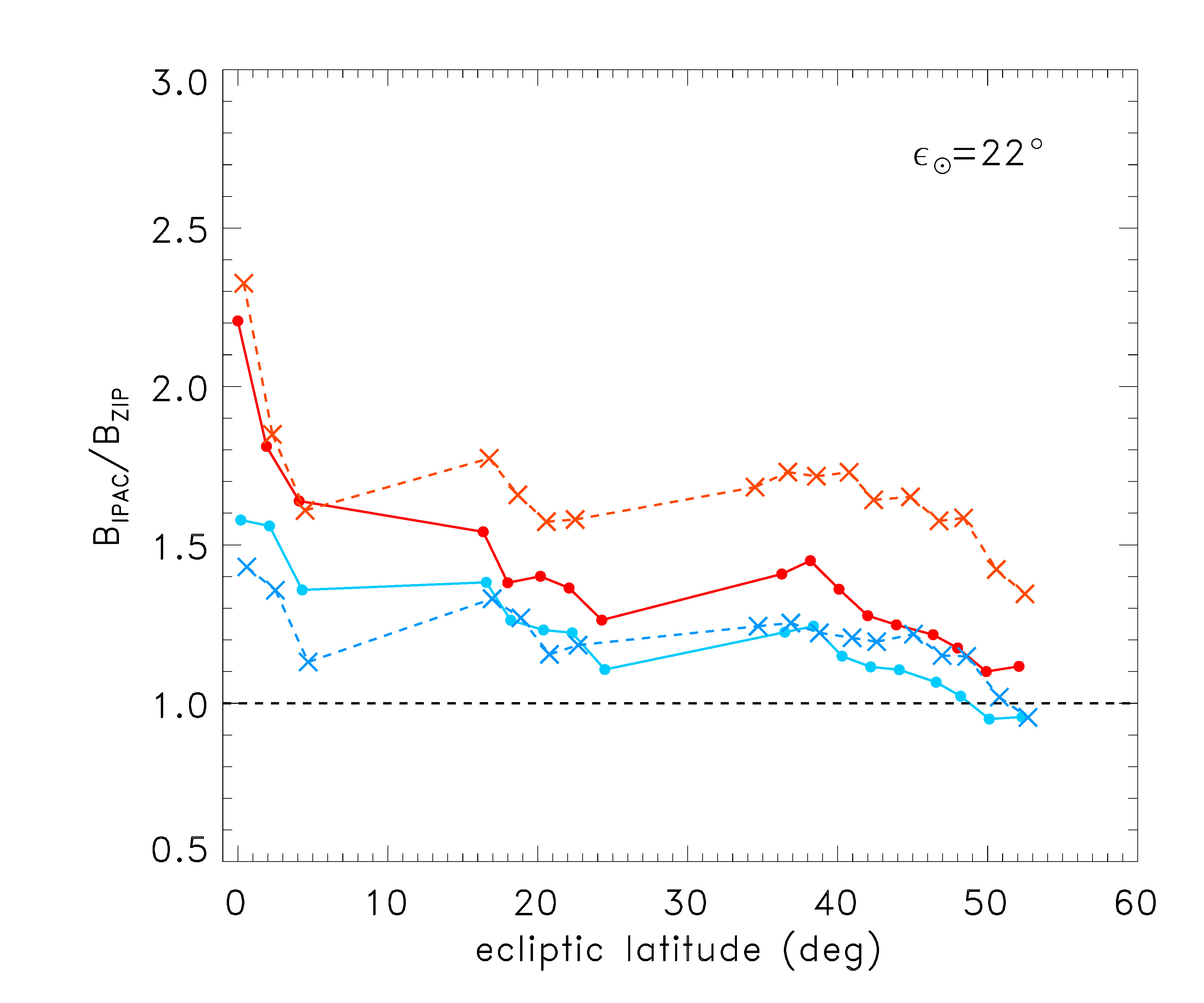}
    \includegraphics[width=0.33\linewidth]{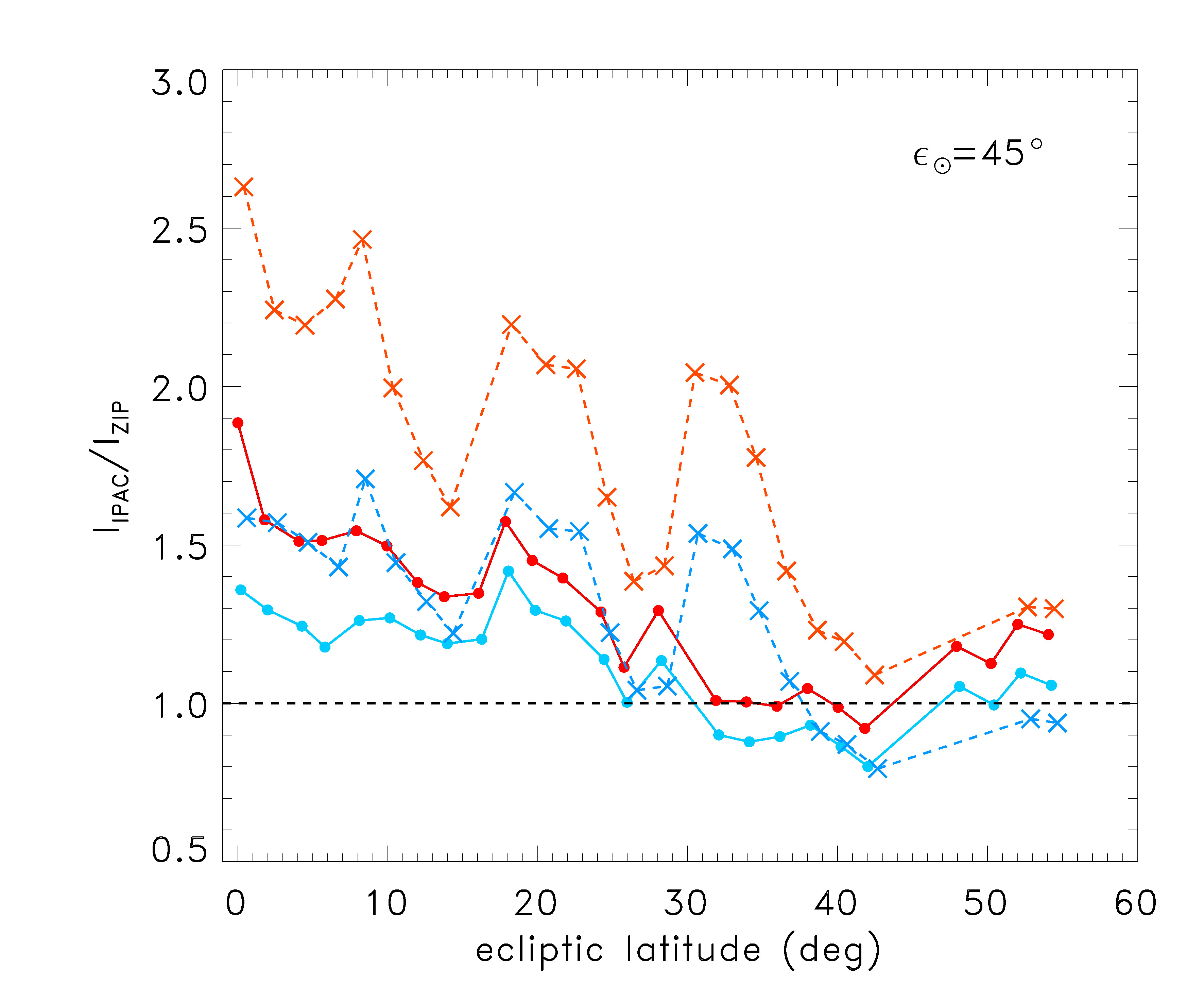}
    }
    \caption{Ratio of model to measured surface brightness values. The arrangement of the figures, and the symbols and the colours are the same as in Fig.~\ref{fig:zipsfb}. }
    \label{fig:zipratio}
\end{figure*}

In \citet{Murdock1985AJ.....90..375M}, ZIP data are presented (1) as a function of solar ecliptic elongation ($\epsilon_\odot$\,=\,$|\lambda-\lambda_\odot|$) when the pointings are in the zodiacal symmetry plane or (2) as a function of ecliptic latitude for specific, fixed $\epsilon_\odot$ values ($\epsilon_\odot$\,=\,22, 45, 60, 90 and 180\degr). For the measurements in the symmetry plane, we adopted a symmetry plane inclination of i\,=\,2\degr, and ascending node longitude of $\Omega$\,=\,78\degr, in agreement with the values used in \citet{Kelsall1998}. 
We note that there is a slight wavelength-dependence of these values which do not change our results notably.
For the fixed $\epsilon_\odot$ cases, we calculated the model predictions of all possible combinations of $\lambda$\,=\,$\lambda_\odot \pm \epsilon_\odot$ and $\pm\beta$.
 In agreement with the goals of the present paper, we used the symmetry plane data and the $\epsilon_\odot$\,=\,22 and 45\degr\ data, and compared them with model predictions \citep[figs.~12, 13, and 14 in][]{Murdock1985AJ.....90..375M}. The results are presented in Figs.~\ref{fig:zipsfb} and \ref{fig:zipratio}.
 The data presented in these figures correspond to the data in figs.~12, 13, and 14 in \citet{Murdock1985AJ.....90..375M}. The results are presented in $I_\lambda$ ($\mathrm{W\,cm^{-2}\,\mu m^{-1}\,sr^{-1}}$) units rather than in $I_\nu$ (MJy\,sr$^{-1}$), as this representation provides a clearer visual separation of the data at the two wavelengths in the figures. (Conversion between the two units is done by I(\mjysr)\,=\,I(\Wcmumsr)$\times k_0(\lambda)$ where $k_0(\lambda)$ are 3.963e11 and 1.457e12 at 10.9 and 20.9\,$\mu$m, respectively, i.e. for example 10$^{-10}$\,\Wcmumsr\ corresponds to 39.6 and 145.7\,\mjysr\ at 10.9 and 20.9\,$\mu$m.).

In the symmetry plane, the deviations from the measured values show no strong dependence on solar elongation (Fig.~\ref{fig:zipratio}, left panel). The Wright model (IPAC V4) overestimates the surface brightness by a factor of $\sim$2.5 in the 20.9\,$\mu$m band and by $\sim$1.5 in the 10.9\,$\mu$m band. The Kelsall model (IPAC V1) performs better, with corresponding factors of $\sim$1.5 and $\sim$1.3 in the two bands.

For cases at constant solar elongation (Fig.~\ref{fig:zipratio}, middle and right panels), a clear dependence on ecliptic latitude is evident. At low ecliptic latitudes, the surface brightness ratios are $\sim$1.5 for the Kelsall model and $\sim$2$-$3 for the Wright model. Both models converge toward unity at higher ecliptic latitudes. This behaviour is expected, as these regions correspond to solar elongations $\gtrsim 60^\circ$, i.e., configurations similar to those directly constrained by COBE/DIRBE observations. Even in this regime, however, the Kelsall model remains consistently closer to unity than the Wright model.

\section{Conclusions \label{sect:conclusions}}

\begin{figure}[ht!]
    \centering
    \includegraphics[width=0.99\linewidth]{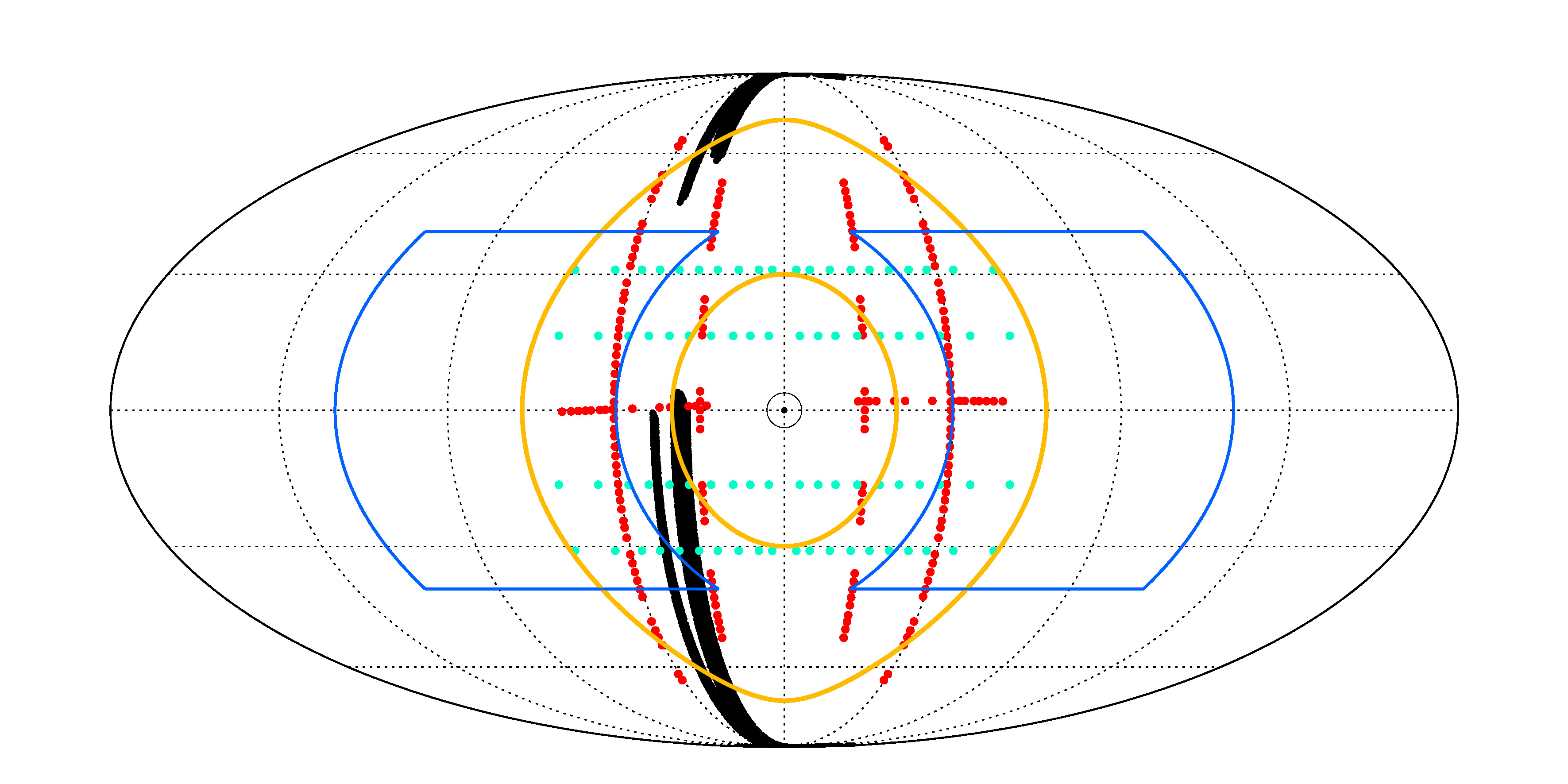}
    \caption{{All-sky map of the location of the  measurements used in this work, in solar ecliptic coordinates ($\lambda-\lambda_{\odot}$, $\beta$), in 
    Mollweide projection (the Sun is in the centre of the map). }
   Black, red and green dots mark MSX, ZIP and Helios data, respectively. Orange and blue curves mark the fields of regard of the NEOMIR and NEO Surveyor missions, respectively.}
    \label{fig:skyloc}
\end{figure}

When evaluating the results, it is important to consider the sky coverage of the individual datasets. Although the MSX data provide the largest number of measurements and cover four wavelength bands, they only sparsely sample the zodiacal symmetry plane ($\beta \approx 0^\circ$) and small solar elongations (Fig.~\ref{fig:skyloc}). The ecliptic-crossing scans are confined to a relatively small region of the sky around $\lambda \approx 100^\circ$ in ecliptic coordinates,  and especially in solar ecliptic longitude of $\lambda-\lambda_{\odot}$\,$\approx$\,330\degr\, as the observations were obtained during a limited time interval of approximately one month. Owing to a different observing strategy, the ZIP data probe the zodiacal symmetry plane as well as cross-ecliptic scans at a range of solar elongations, with partial overlap in sky coverage with the MSX observations. The Helios measurements sample a largely distinct region of the sky and, although they do not probe the symmetry plane, provide observations at $|\beta|$\,=\,16\degr and 31\degr, extending the combined dataset to a broader range of ecliptic longitudes than covered by the MSX and ZIP observations.

\begin{table}[ht!]
\vspace{0.3cm}
\centering
\renewcommand{\arraystretch}{1.0} 
\begin{tabular}{lcccccccccc}
\hline
\rowcolor{gray!20}
 Dataset & Wavelength & V1/obs  & V4/obs  \\
         & ($\mu$m)   &         &         \\
\hline
  Helios & 0.53 & 0.8–1.0   &   1.5–-1.8 \\
 MSX A   & 8.28  & $\sim$1.0 &   1.1–-1.2 \\
 MSX C   & 12.13 & $\sim$0.9 &  0.9--1.1 \\
 MSX D   & 14.65 & 0.9--1.0 & $\sim$1.2 \\
 MSX E   & 21.34 & $\sim$0.8 & $\sim$1.2 \\
 ZIP     & 10.9 & $\sim$1.3  & $\sim$1.5 \\
 ZIP     & 20.9 & $\sim$1.5  & $\sim$2.5 \\
\hline
\end{tabular}
\caption{Summary of general performance of the zodiacal light surface brightness estimator tools. The columns are: Name of the probe/instrument, central wavelength of the photometric band, typical ratio of the model and observed surface brightness values in the Kelsall / IPAC V1 and Wright / IPAC V4 models.}
\label{table:summary}
\end{table}

The overall performance of the zodiacal-emission models in the solar-elongation range $\epsilon \approx 20^\circ$--$60^\circ$ is summarized in Table~\ref{table:summary}. Considering that both models were primarily constrained by observations obtained at $\epsilon \gtrsim 60^\circ$, their performance at moderate elongations ($\sim$40--70\degr) is generally satisfactory. Our analysis shows that the Kelsall model (IPAC V1 implementation) consistently provides a more accurate representation of the observed zodiacal background than the Wright model (IPAC V4). In particular, at wavelengths most relevant for near-Earth asteroid discovery missions ($\sim$8--10\,$\mu$m), the Kelsall model typically reproduces the observed surface brightness to within $\sim$10\% over a broad range of viewing geometries. Even at the smallest solar elongations investigated ($\epsilon \approx 20^\circ$--$30^\circ$), where the uncertainties are largest, the Kelsall model generally outperforms the Wright model.

Nevertheless, the Wright-based model (IPAC~V4) performs better in some wavelength and solar-elongation ranges.  In these cases, the Kelsall model tends to underestimate the zodiacal emission, although typically by no more than $\sim$10\%. This behaviour is evident in the MSX band A and C observations (8.3 and 12.1\,$\mu$m; Fig.~\ref{fig:msxbands}), as well as in the ZIP measurements at $\epsilon_\odot = 45^\circ$ and $\beta \approx 40^\circ$ (Fig.~\ref{fig:zipratio}). However, the MSX observations are confined to a relatively small region of the sky, whereas the ZIP measurements sample a broader range of viewing geometries and indicate a more general tendency of the Wright-based model to overestimate the zodiacal emission.

Overall, these results suggest that implementations based on the Kelsall model, such as IPAC V1 and ZodiPy, provide a more reliable basis for performance estimates and sensitivity assessments of infrared space missions operating at small solar elongations. However, both models exhibit systematic, wavelength-dependent deviations from the observations. At moderate elongations ($\epsilon \approx 40^\circ$--$70^\circ$), discrepancies of order 10\%--20\% are common, while at the smallest elongations ($\epsilon \lesssim 30^\circ$) both models can depart significantly from the measured surface brightness. Although the Kelsall model remains the more accurate of the two, these results indicate that the zodiacal emission at small solar elongations is still not fully constrained by existing observations. 

The combined MSX and ZIP data indicate that the Kelsall model overestimates the zodiacal brightness near the ecliptic plane by approximately 30\%, while elsewhere the typical discrepancy is $\sim$12\%, comparable to or moderately exceeding the calibration uncertainties. Because MSX provides limited coverage near $\beta=0^\circ$, we adopt indicative empirical uncertainties of 30\% for $|\beta|\lesssim20^\circ$ and $\epsilon\lesssim40^\circ$, and 12\% elsewhere; these are not formal confidence limits, given the limited number and sky coverage of the available measurements.

Our analysis also highlights that the zodiacal emission at very small solar elongations ($\epsilon \lesssim 30^\circ$) remains only weakly constrained by existing observations. 
Additional thermal-infrared observations at small solar elongations would therefore be highly valuable, both for refining existing zodiacal-light models and for improving background estimates critical to the design and performance assessment of future infrared space telescopes and near-Earth object discovery missions.

{Although NEO Surveyor is not designed as an absolute diffuse-background photometer, accurate zodiacal-emission predictions remain important for
its performance assessment. The mission will survey at solar elongations down to $\sim45^\circ$ in its NC1 ($4$--$5.2\,\mu{\rm m}$) and NC2 ($6$--$10\,\mu{\rm m}$) channels, where zodiacal emission is expected to be the dominant diffuse source of photon noise. Uncertainty in this background therefore propagates directly into estimates of limiting sensitivity, required integration time, detector dynamic range, survey
completeness, and NEO discovery yield. The available measurements indicate that Kelsall-based predictions are generally reliable to within
approximately $10$--$20\%$ at the elongations most relevant to NEO Surveyor, although larger uncertainties remain near the ecliptic plane
and toward smaller elongations. Whether NEO Surveyor observations themselves can provide sufficiently accurate absolute diffuse-background
measurements to refine these models remains uncertain; nevertheless, improved zodiacal-emission constraints are directly relevant to both
mission planning and the interpretation of its survey performance.}

\begin{acknowledgments}
The research leading to these results has received
funding from the K-138962 grant of the National Research, Development and Innovation Office (NKFIH), Hungary.
JL acknowledge support from the Agencia Estatal de Investigaci\'on del Ministerio de Ciencia e Innovaci\'on (AEI-MCINN) under the grant "Hydrated Minerals and Organic Compounds in Primitive Asteroids 2" with reference PID2024-160618NB-C22. We thank the reviewer for their constructive comments, helpful suggestions, and insightful clarifications on several points.
\end{acknowledgments}

\facilities{MSX, Helios A, Helios B, ZIP}

\bibliography{bib}{}
\bibliographystyle{aasjournalv7}

\newpage

\clearpage

\end{document}